\documentclass[]{aa}

\usepackage[varg]{txfonts}
\usepackage{siunitx}
\usepackage{subcaption}
\usepackage{tabularx}
\usepackage{booktabs} 
\usepackage{graphicx}
\usepackage{verbatim} 
\usepackage{hyperref}
\usepackage[dvipsnames]{xcolor}

\numberwithin{equation}{section} 
\usepackage{natbib}
\bibpunct{(}{)}{;}{a}{}{,} 

\begin{document}
    \title{Distinguishing exoplanet companions from field stars in direct imaging using Gaia astrometry}
\titlerunning{Distinguishing exoplanet companions from field stars in direct imaging using Gaia astrometry}

\author{Philipp Herz, Matthias Samland, Coryn A.L.\ Bailer-Jones\thanks{email: calj@mpia.de}}
\institute{Max-Planck-Institut für Astronomie, Königstuhl 17, 69117 Heidelberg, Germany}
\authorrunning{Herz, Samland, Bailer-Jones}

\date{Submitted 6 Nov.\ 2023, revised 29 Nov.\ 2023, accepted 7 Dec.\ 2023}

\abstract{

Direct imaging searches for exoplanets around stars detect many spurious candidates that are in fact background field stars.
To help distinguish these from genuine companions, multi-epoch astrometry can be used to identify a common proper motion with the host star. Although this is frequently done, many approaches lack an appropriate model for the motions of the background population, or do not use a statistical framework to properly quantify the results.
Here we use Gaia astrometry combined with 2MASS photometry to model the parallax and proper motion distributions of field stars around exoplanet host stars as a function of candidate magnitude. We develop a likelihood-based method that compares the positions of a candidate at multiple epochs with the positions expected under both this field star model and a co-moving companion model. Our method propagates the covariances in the Gaia astrometry and the candidate positions. True companions are assumed to have long periods compared to the observational baseline, so we currently neglect orbital motion.
We apply our method to a sample of 23 host stars with 263 candidates identified in the B-Star Exoplanet Abundance Study (BEAST) survey on VLT/SPHERE.
We identify seven candidates in which the odds ratio favours the co-moving companion model by a factor of 100 or more.
Most of these detections are based on only two or three epochs separated by less than three years, so further epochs should be obtained to reassess the companion probabilities.
Our method is publicly available as an open-source python package from \href{https://github.com/herzphi/compass}{GitHub} to use with any data.

\vspace*{1em}
}

\keywords{Exoplanets - Astrometric motion - Direct imaging - open source software}
\maketitle

\begin{figure}
    \centering
    \includegraphics[width=\linewidth]{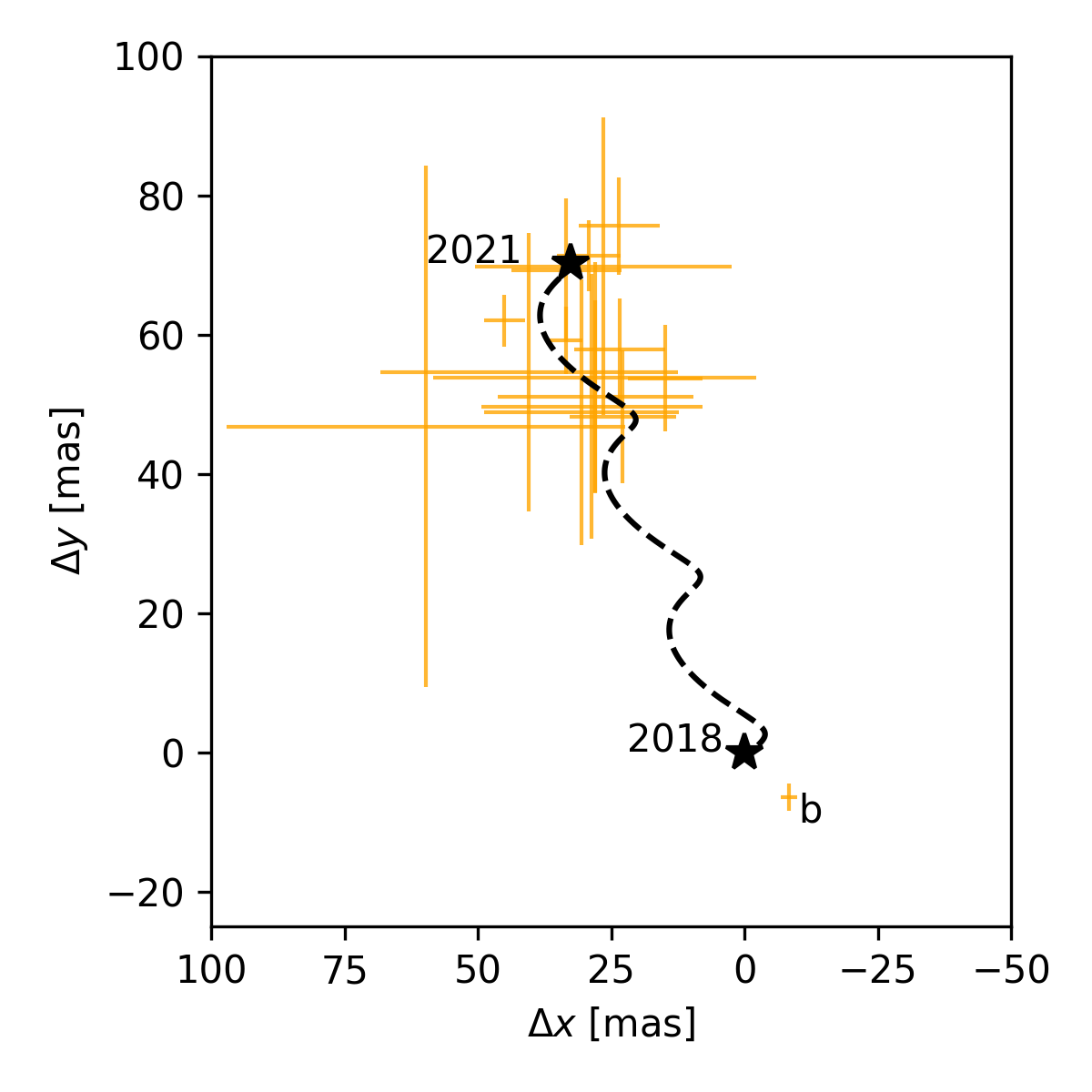}
    \caption{Change in position of exoplanet candidates (orange crosses) relative to the star $\mu^2$ Sco between two measurement epochs. A co-moving source should be close to the origin (labeled "2018"). A background source with zero proper motion will move according to the dashed curve (a reflection of the host star's parallax and proper motion) ending in the black star labeled "2021". The motion of $\mu^2$ Sco~b is distinct from the cloud of background stars in the field that are (through this plot) deemed not to be exoplanets. Figure adapted from \cite{Squicciarini_2022}.}
    \label{fig:astro_motion}
\end{figure}

\section{Introduction}
Young exoplanets with a favourable brightness and separation to their host star can be directly imaged. However, such exoplanets can be confused with more-distant background stars that happen to lie in the line-of-sight. A common way to distinguish these scenarios is to observe both the host star and candidate over time to look for a common proper motion and/or parallax.
An example of doing this is shown in Fig.~\ref{fig:astro_motion} (data from \citealt{Janson_2021} and published in \citealt{Squicciarini_2022}).
The positions of the host star and a number of exoplanet candidates were measured at two epochs, 2018 and 2021. The orange crosses show the measured change in position (relative to the host star) of candidates between these two epochs.
The host star has some proper motion and parallax between 2018 and 2021: the black dashed line shows how objects with zero proper motion and parallax would move over this time period (as our view is centred on the host star).
Those orange points clustered near the black point labelled 2021 are therefore consistent with being distant background stars. The candidate labelled ``b'', on the other hand, has a motion more consistent with the host star and so is more likely to be a true companion.

To make this procedure quantitative we must take into account the measurement uncertainties and any covariance between them. Background stars do not have zero parallax and proper motion, so we need a proper model for their motions too. There may also be more than two epochs, so we will want to take into account all of the data within a single assessment. In the literature these confounding factors are usually not considered \citep[e.g.][]{Lagrange2010, Lafreniere2011, Carson2013, Kuzuhara2013, DeRosa2015, Konopacky2016, Chauvin2017, Keppler2018, Bohn2021, Janson_bCen_2021, Squicciarini_2022, Franson2023, Mesa2023, Chomez2023}.
Such studies often compare the second epoch position with that expected for a stationary background star. If they deviate significantly the null hypothesis of being a background star is rejected and some alternative model, i.e.\ a companion, is implicitly accepted. This classical hypothesis testing approach does not, however, assess whether the data might be even more unlikely under the companion model.

The goal of this paper is to put this astrometric confirmation approach on a solid statistical footing.
We develop a model to evaluate whether the multi-epoch motion of the candidate is more likely to be a co-moving exoplanet or a coincidental field star.
Our model is based on the proper motion and parallax distributions of field stars in the same area of the sky as the candidate, and with similar magnitudes to the candidate being tested. 
Using the odds ratio, we compare this background model with a model in which the candidate is co-moving with the host star.
We apply our method to multi-epoch measurements of candidates observed in the B-Star Exoplanet Abundance Study (BEAST) survey \citep{Janson_2021}, which has an abundance of candidate objects due to observing close to the Galactic plane.
A python package implementing our method is accessible via GitHub\footnote{\href{https://www.github.com/herzphi/compass}{github.com/herzphi/compass}}.
    \section{Methods}\label{sec:Analysis}
We develop a probabilistic method that compares the likelihoods of position measurements of an exoplanet candidate under two models: the first assumes the candidate is a co-moving companion, the second assumes it is a field star. The former uses the proper motion and parallax of the exoplanet candidate's host star, the latter additionally uses a magnitude-dependent fit to the parallax and proper motion distributions of field stars close to the host star's line-of-sight. Our method can use an arbitrary number of epochs of astrometric observations to derive the likelihoods. We also implement a special case of the general method that neglects parallax. This is less realistic but is easier to visualize. The field star astrometric model is currently built on data from the Gaia \citep{2023A&A...674A...1G} and 2MASS \citep{2006AJ....131.1163S} surveys, but could use other present or future survey.

\subsection{Positional model}\label{sec:positional_model}

Consider the position of a candidate object relative to the host star as a function of time. Assuming that the candidate does not move relative to the star, then in a geocentric coordinate system this can be described as a linear motion with a superimposed parallactic motion. In a Cartesian plane projection we write this as
\begin{equation}\label{eq:rel_position_Nepochs}
    \begin{split}
        \Delta {x}_i' &= \left( {x}_{a,0}'-{x}_{\star,0}' \right) + \left( \mu_{a, x}-\mu_{\star, x} \right)t_i + \left( \varpi_{a}-\varpi_{\star} \right){s_x}(t_i)\\
        \Delta {y}_i' &= \left( {y}_{a,0}'-{y}_{\star,0}' \right) + \left( \mu_{a, y}-\mu_{\star, y} \right)t_i + \left( \varpi_{a}-\varpi_{\star} \right){s_y}(t_i)\\
    \end{split}
\end{equation}
where $\mu$ is the proper motion and $\varpi$ is the parallax. Subscript $a$ refers to the candidate object, subscript $\star$ to the host star, subscript $0$ to the true position of the candidate at a reference epoch, and subscript $i$ to the $i$-th epoch relative to the reference epoch. The functions $s_x$ and $s_y$ are periodic phase factors for the parallax motion given by the orbital motion of the Earth around the Sun at epoch $i$. The primed variables denote the true position of the candidate, while the unprimed variables denote the measured position of the candidate. In this paper ``true positions'' denote those we would obtain in the absence of noise in our positional measurements at the epochs, but still based on some externally-provided values of parallax and proper motion. Those external values {\em are} considered to be noisy, and their covariances will be taken into account.

Equation~\ref{eq:rel_position_Nepochs} can be generalized to $N$ measurements of the candidate's position over time $(\Delta\boldsymbol{x}, \Delta\boldsymbol{y}) = \{\Delta x_i, \Delta y_i\}$ at times $\boldsymbol{t} = \{t_i\}$ with $i\in[1,N]$, where the bold font indicates a vector of 
measurements.
The first position measurement defines the true position of the candidate in the first epoch $(\Delta x_1, \Delta y_1) = (\Delta x_1', \Delta y_1')$ without loss of generality.

\subsection{Overall probabilistic model}\label{sec:overall_prob_model}
Given the measured positions of the exoplanet candidate over time, we compute the likelihood of the data under two models. 
The first model, denoted $M_{c}$, assumes the exoplanet candidate is a co-moving companion, sharing the same proper motion and parallax as the host star. We make the simplifying assumption that the orbital period is long compared to the observational baseline and so neglect the candidate's orbital motion. This may not be justified in all cases, which we elaborate on in Section~\ref{sec:discussion}. The second model, denoted $M_b$, assumes the candidate to be a background object\footnote{We use the term ``background objects'' and ``field stars'' interchangeably, as most field stars are more distant than the host star.} with a proper motion and parallax distribution constructed from a set of background stars in a narrow field-of-view around the host star (described in Section~\ref{sec:FSmodel}). We then compare these two models via the odds (likelihood) ratio 
\begin{equation}\label{eq:pm_oddsratio}
    r_{c, b} = \frac{P(\Delta\boldsymbol{x}, \Delta\boldsymbol{y}\mid M_{c})}{P(\Delta\boldsymbol{x}, \Delta\boldsymbol{y}\mid M_{b})}\ .
\end{equation}
which indicates which model favours the data more. This does not give the posterior probability of the model given the data, however, for which we would first need to establish model prior probabilities.

Let us first consider the denominator in Eq.~\ref{eq:pm_oddsratio}, the likelihood under the background model.
This can be computed as a combination of two probability density functions (PDFs). The first of these is the probability of the noisy position measurements given the true positions,
$P(\Delta\boldsymbol{x}, \Delta\boldsymbol{y}\mid \Delta\boldsymbol{x}', \Delta\boldsymbol{y}')$, which reflects the noise in the determination of the centroid of a point-spread-function on the detector. We assume this to be a Gaussian 
with mean $(\Delta\boldsymbol{x}', \Delta\boldsymbol{y}')$ and a covariance matrix $\Lambda$ that reflects the accuracy of, and correlations in, the measurements.
The second term is $P(\Delta \boldsymbol{x}', \Delta \boldsymbol{y}' \mid \boldsymbol{t}, M_{b})$. 
This represents the spread in possible ``true'' positions of a background star at specific times arising from the spread in the proper motion and parallax of the background star population.
We assume this term to be a Gaussian distribution with a mean given by the right side of Eq.~\ref{eq:rel_position_Nepochs}. The terms in that equation, as well as the covariance matrix of the Gaussian (which we denote $\Lambda'$), come from our fit to the background star population (Section~\ref{sec:FSmodel}).
These two PDFs we then combine via a marginalization to give the required likelihood
 \begin{equation}\label{eq:PxN_rel}
    \begin{split}
        P\left( \Delta\boldsymbol{x},\Delta\boldsymbol{y} \mid \boldsymbol{t}, M_b \right) &= \int_{\Delta\boldsymbol{x}', \Delta\boldsymbol{y}'} P\left( \Delta\boldsymbol{x}, \Delta\boldsymbol{y}\mid \Delta\boldsymbol{x}', \Delta\boldsymbol{y}'\right)
        \,\times \\ &\hspace*{2.5em} P\left(\Delta\boldsymbol{x}', \Delta\boldsymbol{y}' \mid \boldsymbol{t}, M_b \right) d\Delta\boldsymbol{x}'d\Delta\boldsymbol{y}' \ .
    \end{split}
\end{equation}
This shows that the background model likelihood is a convolution of the two PDFs.
Given that these are both Gaussian, the result of the convolution is also a Gaussian, with mean $(\Delta\boldsymbol{x}', \Delta\boldsymbol{y}')$ and covariance matrix $(\Lambda+\Lambda')$.
More details are provided in Appendix~\ref{sec:2Ndmodel}.

We take the same approach to computing 
the likelihood of the data under the 
co-moving companion model $M_{c}$, the
numerator in Eq.~\ref{eq:pm_oddsratio}.
Here, however, the relative position of the exoplanet candidate is assumed to be constant, meaning that the relative position measurements have zero variance and so 
$P(\Delta \boldsymbol{x}', \Delta \boldsymbol{y}' \mid \boldsymbol{t}, M_{c})$ is simply a delta function. 
The convolution in Eq.~\ref{eq:PxN_rel} is then trivial, resulting in 
$P(\Delta \boldsymbol{x}, \Delta \boldsymbol{y} \mid \boldsymbol{t}, M_{c})$ being a Gaussian with mean $(\Delta\boldsymbol{x}', \Delta\boldsymbol{y}')$ and covariance matrix $\Lambda$.

Now that we have expressions for both likelihoods in terms of the measured relative positions and their covariances, and in terms of the parallax and proper motion distribution of the background model, we can compute the odds ratio in Eq.~\ref{eq:pm_oddsratio} and decide which model better explains the data.

The above general method we refer to as the proper motion and parallax covariance method. 
If the parallax is negligible, or if the observations are separated by almost exactly a year, then we can set the parallax in  Eq.~\ref{eq:rel_position_Nepochs} to zero.
If we also have only two epochs, then instead of using the two position measurements directly, we can convert them into a single proper motion.
We call this the "proper motion-only" method.
In this special case, the ``true'' proper motion $\boldsymbol{\Delta \mu}'=(\Delta \mu'_x, \Delta \mu'_y)$ relative to the host star can be written
\begin{equation}
        \Delta \mu'_x = \frac{\Delta{x}_2'-\Delta{x}_1'}{t_2-t_1}
\end{equation}
which we compare to the measured relative proper motions from the data (same equation without the accents). 
The likelihoods for the two models are then
\begin{equation}\label{eq:pm_likleihood_truecompanion}
     P\left(\boldsymbol{\Delta \mu}\mid M_{c}\right) = \mathcal{N} \left(\begin{bmatrix}
         0 \\
         0
     \end{bmatrix}, 
     \begin{bmatrix}
         \text{Var}(\Delta \mu_{x}) & \text{Cov}(\Delta \mu_{x},\Delta \mu_{y}) \\
         \text{Cov}(\Delta \mu_{x},\Delta \mu_{y}) & \text{Var}(\Delta \mu_{y})
     \end{bmatrix}\right)
 \end{equation}
 and 
 \begin{equation} \label{eq:28}
         P\left(\boldsymbol{\Delta \mu}\mid M_b\right) = \int P\left(\boldsymbol{\Delta \mu}\mid\boldsymbol{\Delta \mu}'\right) P\left(\boldsymbol{\Delta \mu}'\mid M_b\right) d\boldsymbol{\Delta \mu}'.
\end{equation}
Both equations are Gaussian distributions. In the co-moving companion model the likelihood is centered at zero with a covariance matrix given by the uncertainties in position measurements. In the background model the likelihood is centered at the proper motion $\boldsymbol{\Delta \mu}'$ of the background star distribution relative to the host star with a covariance matrix given by the convolution of the proper motions of this background star population with the measurement uncertainties.

A more complete derivation of both the full model and this simpler (but less representative) proper motion-only model is given in Appendix~\ref{sec:2Ndmodel}.\footnote{More details can be found at \url{http://mpia.de/homes/calj/astrometric_companion_model.pdf}.}

\subsection{Proper motion and parallax distribution model for field stars}\label{sec:FSmodel}
In order to assess whether a candidate is astrometrically consistent with a given field star population (i.e.\ to evaluate the likelihood $P\left( \Delta\boldsymbol{x},\Delta\boldsymbol{y} \mid \boldsymbol{t}, M_b \right)$), we have to create a model for the parallax and proper motion distributions of the population. 
Ideally this model would be a function of the true distance and velocity of the star, but we of course do not have this information for an arbitrary candidate. However, we know that distance and velocity -- and therefore parallax and proper motion -- depend on the measurable direction in the Galaxy (e.g.\ \citealt{2023arXiv231100374B}). We can also condition our model on other relevant measurements, the most obvious being magnitude, as this contains some information about distance and stellar population. Ideally we would also use colour (as a crude proxy of mass and age and thus having velocity dependence), but this is often not available for many exoplanet surveys.

\begin{figure}
    \includegraphics[width=\linewidth]{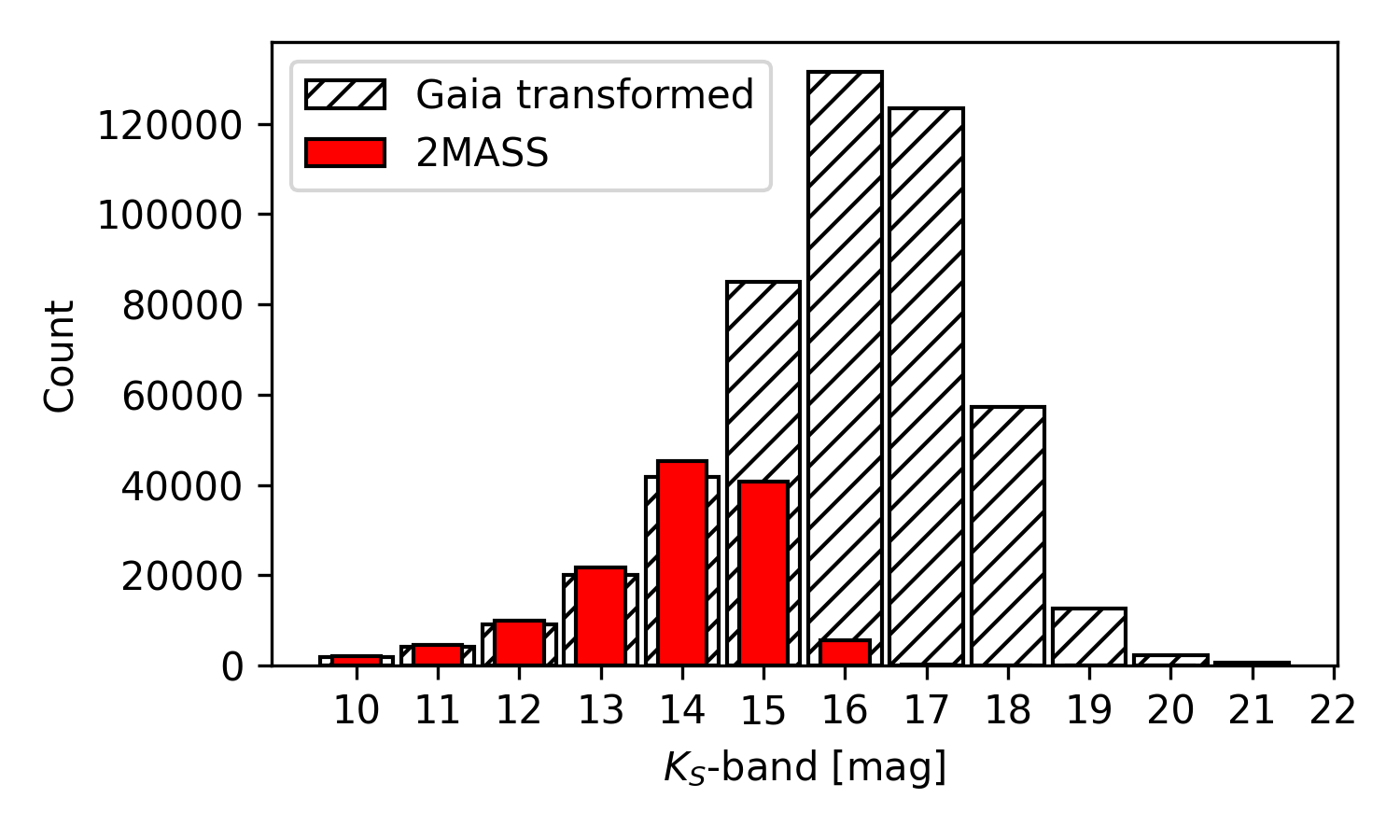}
    \caption{$K_S$-band magnitude distribution of stellar objects in Gaia DR3 in the area of the sky within $0.3^{\circ}$ of the $23$ BEAST stars. Those in red have measured 2MASS $K_S$-band photometry. Those in black stripes have $K_S$ predicted by a colour transformation.}
    \label{fig:ks_band_all_BEAST}
\end{figure}

We build our model empirically using Gaia DR3 \citep{2023A&A...674A...1G}. We select stars that are near to the candidate in Galactic coordinates and that have similar brightness. The latter can be hard to achieve because Gaia observes in the optical and is not as deep as infrared exoplanet surveys. To address this we use a positional cross-match of Gaia with 2MASS \citep{2006AJ....131.1163S}
to assign infrared magnitudes to Gaia stars where possible. For those Gaia objects that do not have a 2MASS counterpart, we assign synthetic $K_S$ magnitudes using the Gaia colour--colour transformation based on the $G$-band magnitude and $G_{\rm BP}-G_{\rm RP}$ colour from \cite{Riello_2021}. 
While this transformation is not perfect,
it significantly increases our field star sample size. Figure~\ref{fig:ks_band_all_BEAST} shows in red all bright field stars covered by 2MASS within  $0.3^{\circ}$ of the 23 target stars of the BEAST. The distribution of stars for which we calculated synthetic $K_S$-Band magnitudes via the colour transformations are shown with black stripes.

Using our sample, we then build a smooth model of the astrometric distributions as a function of magnitude. This will allow us to evaluate (if necessary by extrapolation) the astrometric model at the candidate's brightness.
To build the model we first group the data into magnitude bins, then fit a two-dimensional Gaussian distribution to the proper motions and parallaxes in each bin. 
\begin{figure}
\includegraphics[width=\linewidth]{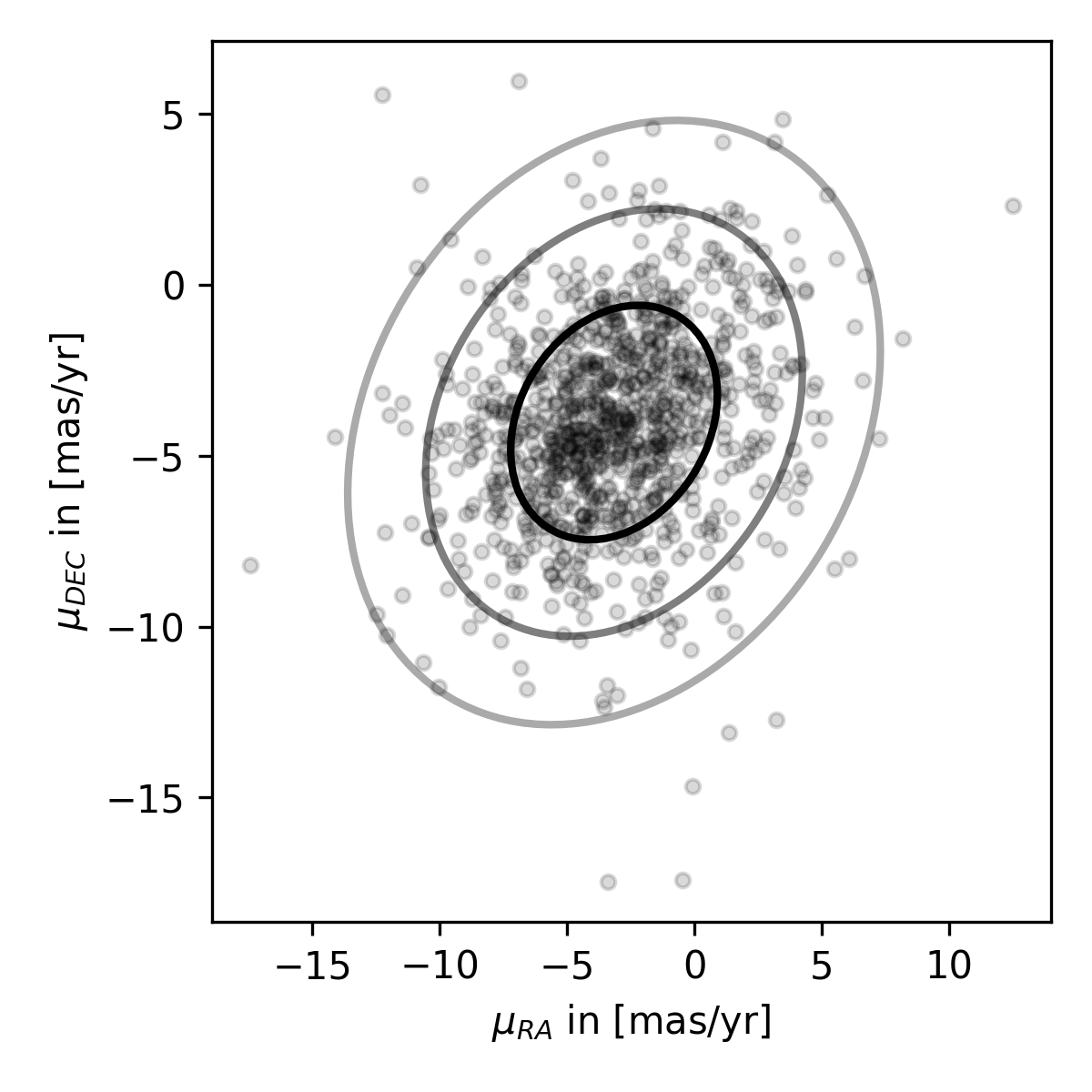}
    \caption{Proper motion distribution of stellar objects with $K_S$-magnitude between $18$ and $19$ in a $0.3^\circ$ sky area around HIP~82545 ($\mu^2$ Sco). The elliptical contours are the boundaries that encompass $50\%$, $90\%$ and $99\%$ of the stellar objects.}
    \label{fig:MAGBIN}
\end{figure}
An example of one magnitude bin is shown in Fig.~\ref{fig:MAGBIN}.
\begin{figure}
\includegraphics[width=\linewidth]{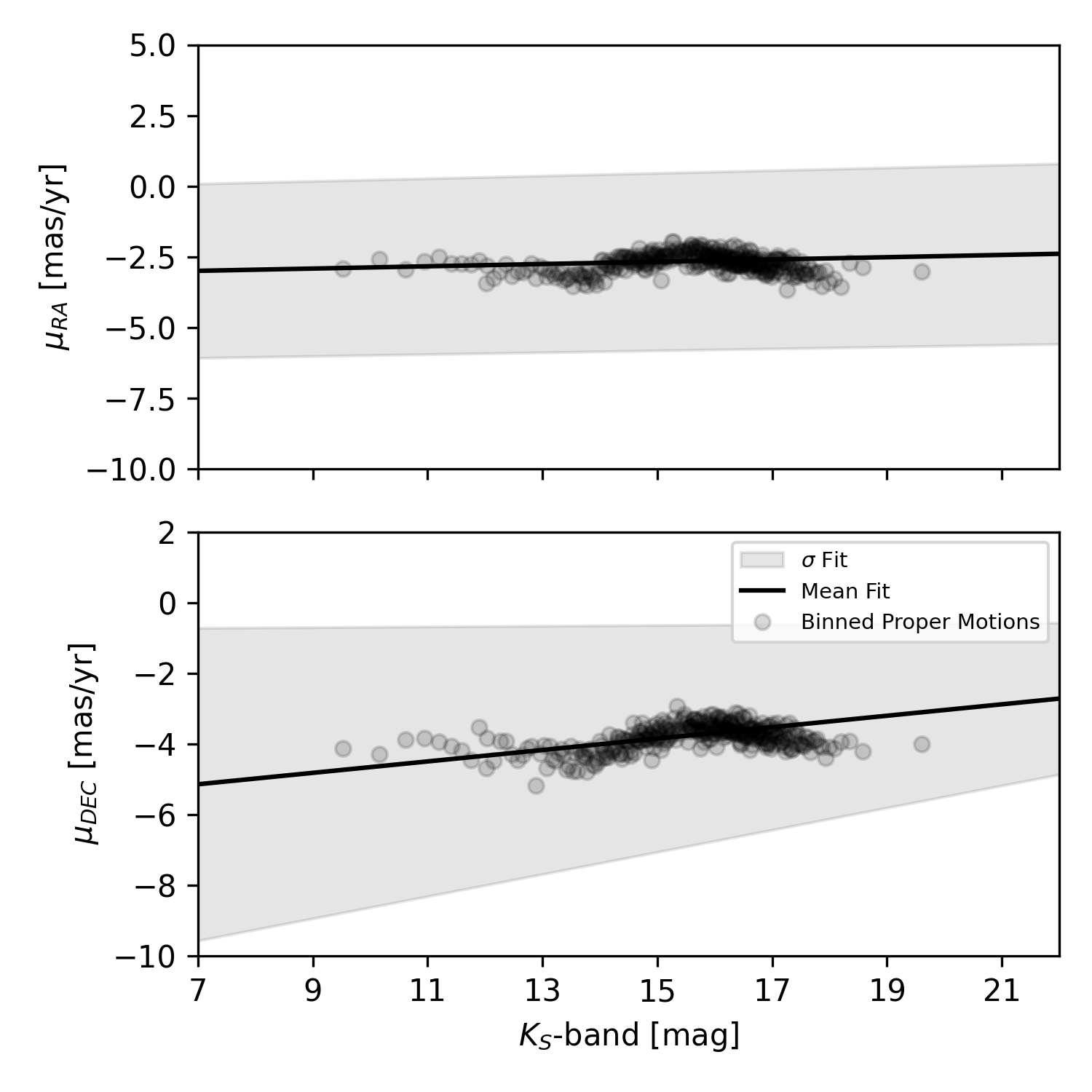}
    \caption{The variation of the mean of the proper motion distribution in field stars as function on stellar magnitude. Each bin includes 200 stellar objects in a $0.3^\circ$ sky area around HIP~82545 ($\mu^2$ Sco). Each point is the mean value of the 2D Gaussian fit in each magnitude bin. Only those points lying within the 10th--90th percentile range of magnitudes are used for the linear fit.}
    \label{fig:meandist_fit}
\end{figure}
We then fit a linear function to the mean values of the 2D Gaussian as a function of magnitude (Fig.~\ref{fig:meandist_fit}), excluding points that lie outside the 10th to 90th percentile range of the magnitudes to achieve a more robust fit. This fit smooths out the variations and allows us to extrapolate the model to fainter candidates than are in Gaia.
\begin{figure}
\captionsetup{width=.9\linewidth}
\includegraphics[width=\linewidth]{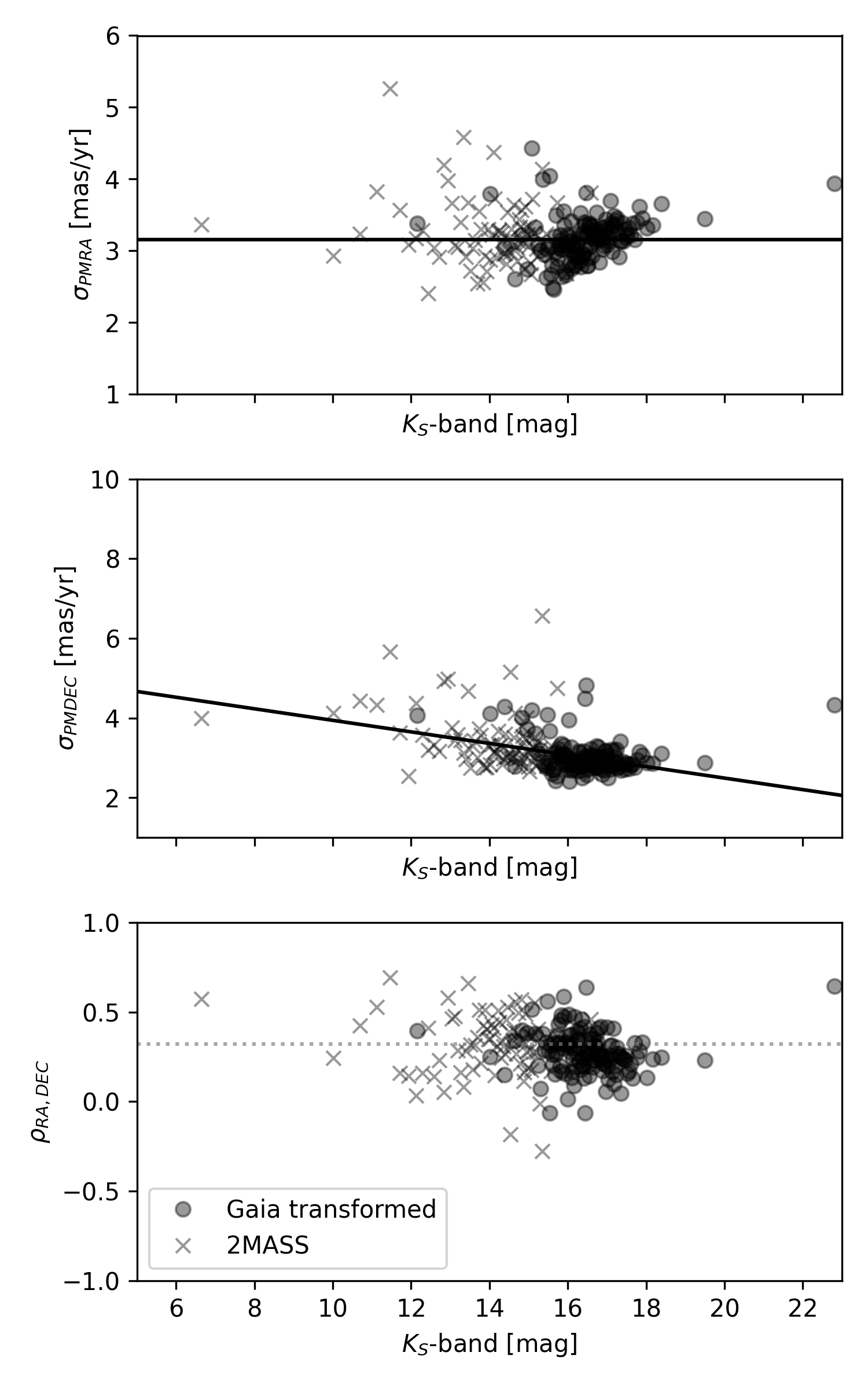}
    \caption{The variation of the standard deviation and correlation of the proper motion distributions for the field stars around HIP~82545 ($\mu^2$ Sco) as a function of magnitude. Each point comes from a 2D Gaussian fit over a narrow magnitude bin (e.g.\ as shown in Fig.~\ref{fig:MAGBIN}).
    Each bin includes 200 stars in a $0.3^\circ$ sky area around the target star. 
    Crosses denote fits using stars with 2MASS magnitudes, and circles denote those with magnitudes computed from the Gaia colour transformation.}
    \label{fig:sigma_rho_fit}
\end{figure}
The standard deviations as a function of magnitude are fit as an exponential
\begin{equation}\label{eq:sigma_min_pm}
    \sigma_\mathrm{fit}\left(m\right) = \sigma_\mathrm{min} + a \exp\bigl(-b(m-m_0)\bigl) 
\end{equation}
with $a>0$. This fit has 
a minimum value of $\sigma_\mathrm{min}$, a fixed constant. The magnitude $m_0$ is the mean of the observed magnitudes. 
An example of this fit is shown in Fig.~\ref{fig:sigma_rho_fit}.
A linear fit is used instead if the exponential fit residuals are larger than those of a linear fit. In making this linear fit we prevent $\sigma_\mathrm{fit}\left(m\right)$ from becoming smaller than 1\,\si{mas/yr}.

We did not identify any definitive and consistent trend in the correlation between parallax and proper motion. We therefore opted to fit a constant value instead.

\subsection{Verification of the method using simulations}

A verification of our model would ideally be based on a large set of real data in which the true nature of each candidate is known, but this is not available.
Here we use simulations to assess the reliability of the odds ratio. We simulate 1000 trajectories of a co-moving companion and 1000 trajectories of background model objects.

Starting with the real star $\mu^2$~Sco, we construct a background model for proper motion and parallax. 
The initial position of the candidate relative to $\mu^2$~Sco is randomly generated, and the position at subsequent epochs (one year apart) is given by
\begin{equation}\label{eq:sim_transformation}
x_{i+1} = x_{i} + \left(\mu_x + \mathcal{N}\left(0,\sigma=3 \si{mas/yr}\right)\right)\Delta t 
\end{equation}
and similarly for $y$.
This equation represents either a co-moving companion with a zero (relative) proper motion ($\mu_x=\mu_y=0$), or a relative proper motion of the background model to $\mu^2$~Sco ($\mu_x = \mu_{RA}(m)-\mu_{x, \mu^2~Sco}$, where $\mu_{RA}(m)$ is a fixed value defined by the brightness of the candidate), both with a normal distribution $\mathcal{N}$ to simulate the propagation of the proper motion uncertainty. Here we choose a standard deviation of $3\si{mas/yr}$.
\begin{figure}
    \centering
\includegraphics[width=0.49\textwidth]{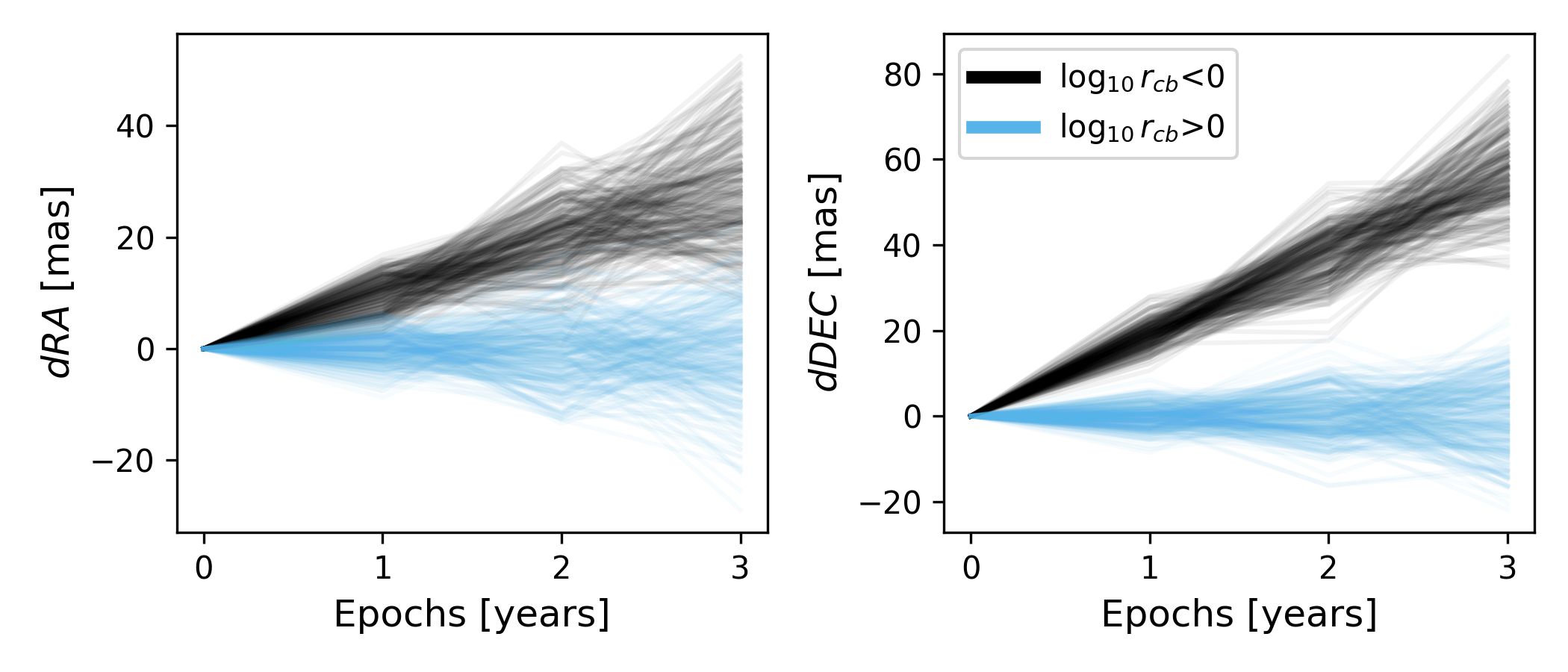}
    \caption{Simulated test of our method by propagating the positions of 2000 simulated candidates to $\mu^2$~Sco over four epochs. Half are propagated according to the co-moving model, which just adds zero mean noise at each epoch. The other half are propagated according to the proper motion of background stars, plus noise. The colours denote the logarithmic odds ratio that our method computes for each trajectory. All 2000 are correctly identified.  }\label{fig:relative_motion_simcand}
\end{figure}
We simulate 1000 paths the candidate could take under the companion model $M_c$ using Eq.\ref{eq:sim_transformation} with $\mu=0$ and another 1000 paths under the background object model $M_b$ using a single fixed proper motion a background object with a similar brightness of the candidate would have. The spread seen in Fig.\ref{fig:relative_motion_simcand} thus arises from the noise we add at each epoch. Our method distinguishes between those paths via a logarithmic odds ratios greater than zero (companion model favoured, show in blue) and smaller than zero (background model favoured, shown in black).
We find that all 2000 trajectories are correctly identified as coming from the model from which they were generated.

    \section{Results}\label{sec:results}
\begin{figure}
    \centering
\includegraphics[width=0.90\linewidth]{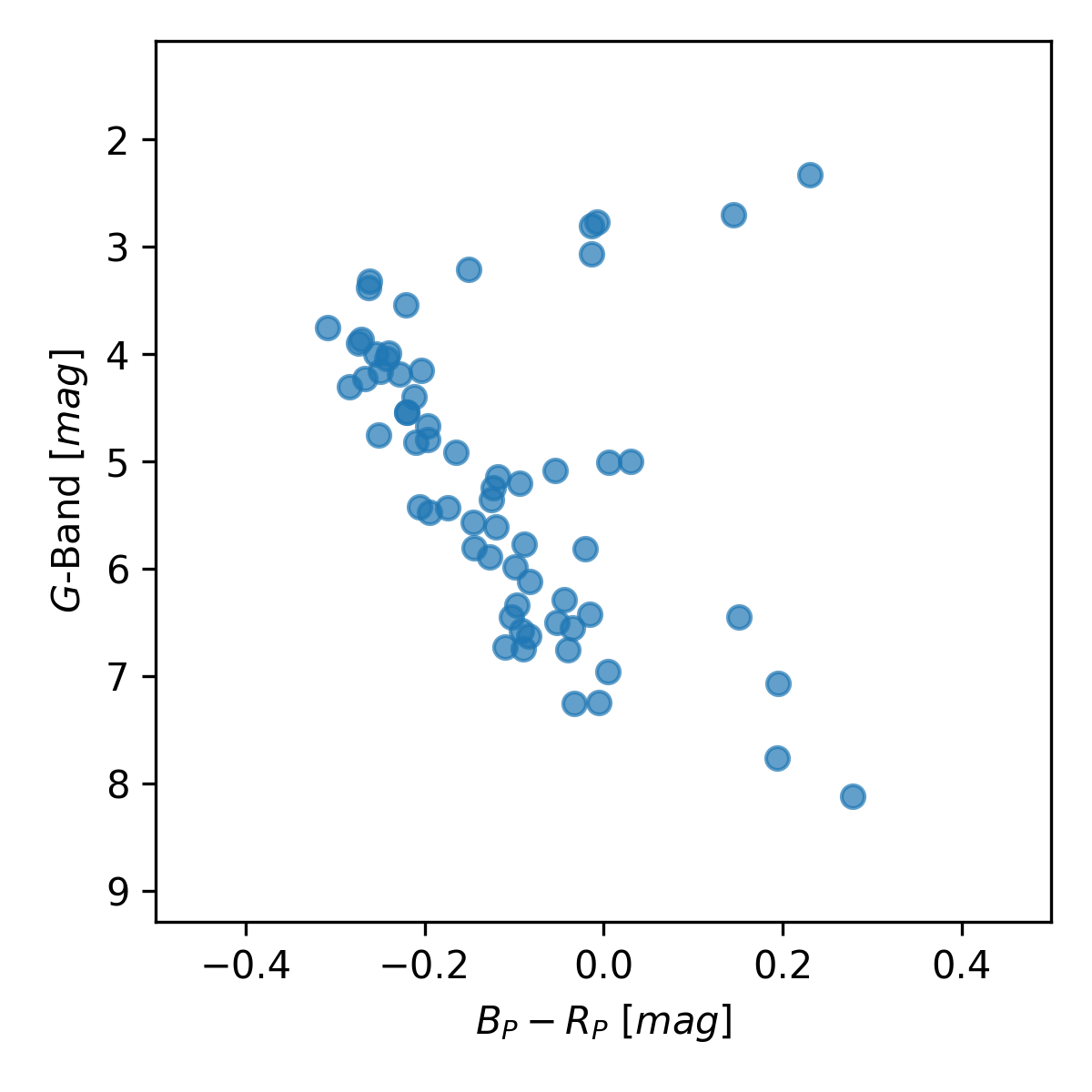}
    \caption{Colour--magnitude diagram of all BEAST stars, except for $\eta$~Cen at $(B_{\rm P}-R_{\rm P}, G)=(4.75, 2.25)$\,mag.}
    \label{fig:cmb_beast}
\end{figure}

The B-star Exoplanet Abundance Study \citep{Janson_2021} employs the VLT/SPHERE \cite{2019A&A...631A.155B} extreme adaptive optics instrument to conduct a direct imaging survey of $85$ B-type stars in the young (5--20~$\si{Myr}$) and relatively nearby (120--150~$\si{pc}$) Scorpius-Centaurus (Sco-Cen) region (Fig.~\ref{fig:cmb_beast}). The targets have similar distance and $G$-Band magnitude, but vary in colour. By targeting young B-type stars, their exoplanets should be relatively bright in the near-infrared. Before this survey, B-stars had not been systematically surveyed for exoplanets. 
Radial velocity surveys, which are more sensitive to close-in planets, reveal comparatively few planets around massive stars \citep{Reffert2015}. Thus BEAST addresses the question of whether massive stars can form massive planets at larger separations. Certain formation scenarios, like disk instability, might primarily take place in the outermost areas of extensive disks surrounding massive stars \citep{Helled2014}.
Sco-Cen is located close to the Galactic plane, so observations often include many spurious candidates in the field-of-view. This makes the survey a prime target for testing our method of distinguishing bound companions from faint field stars.

\begin{figure}
    \centering
    \includegraphics[width=\linewidth]{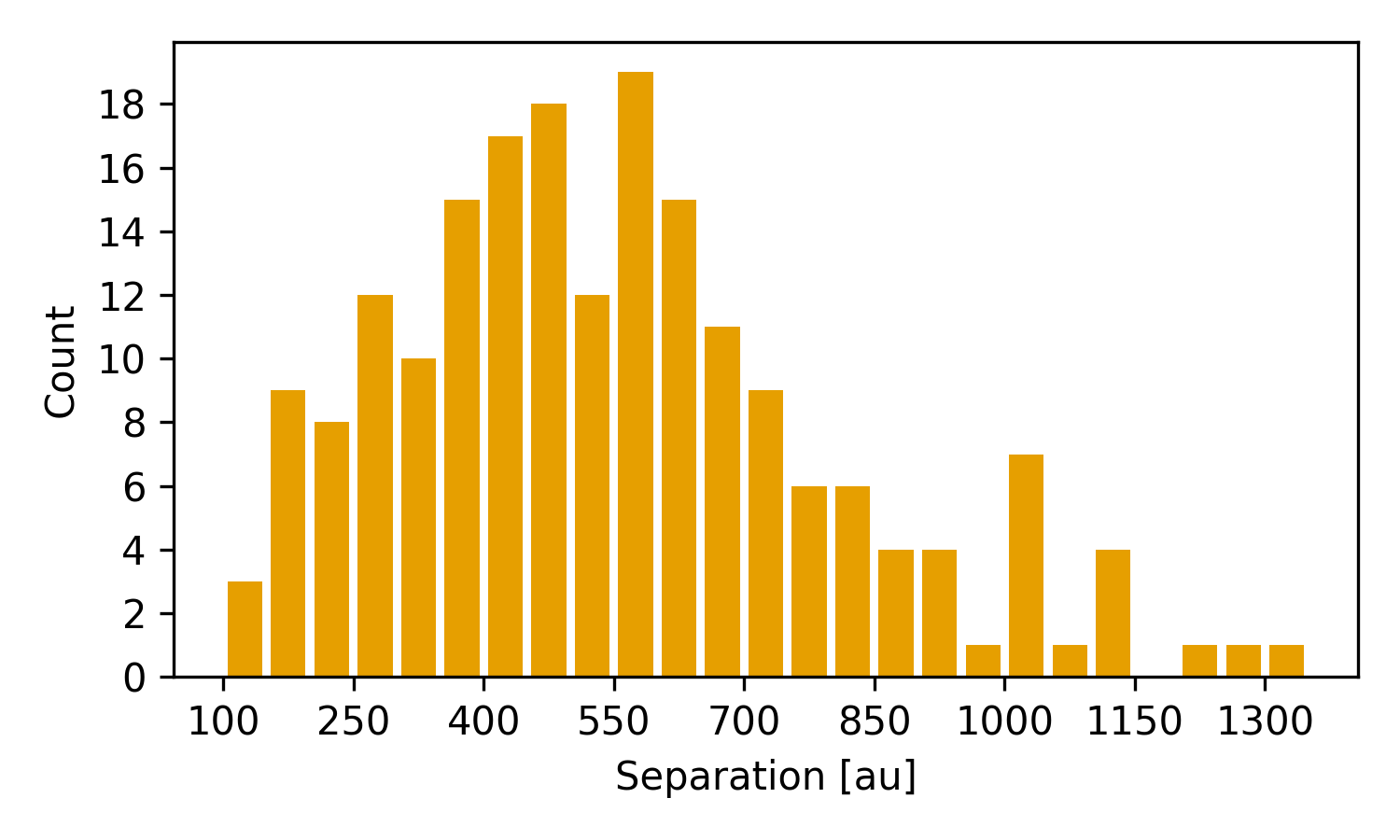}
    \caption{Distribution of the separations of all candidate from all targets identified in BEAST. This assumes the candidates are at the same distance as their respective target stars.}
    \label{fig:sep_beast_candidates}
\end{figure}
Of the $85$ stars in BEAST, $23$ have candidates with at least two epochs of observation (as of 2022-04-01). These 23 stars have in total 263 candidates. The projected separations between the candidates and their host stars range from $49\,\si{au}$ to $1457\,\si{au}$ (Fig.~\ref{fig:sep_beast_candidates}).

\begin{figure}
    \centering
\includegraphics[width=\linewidth]{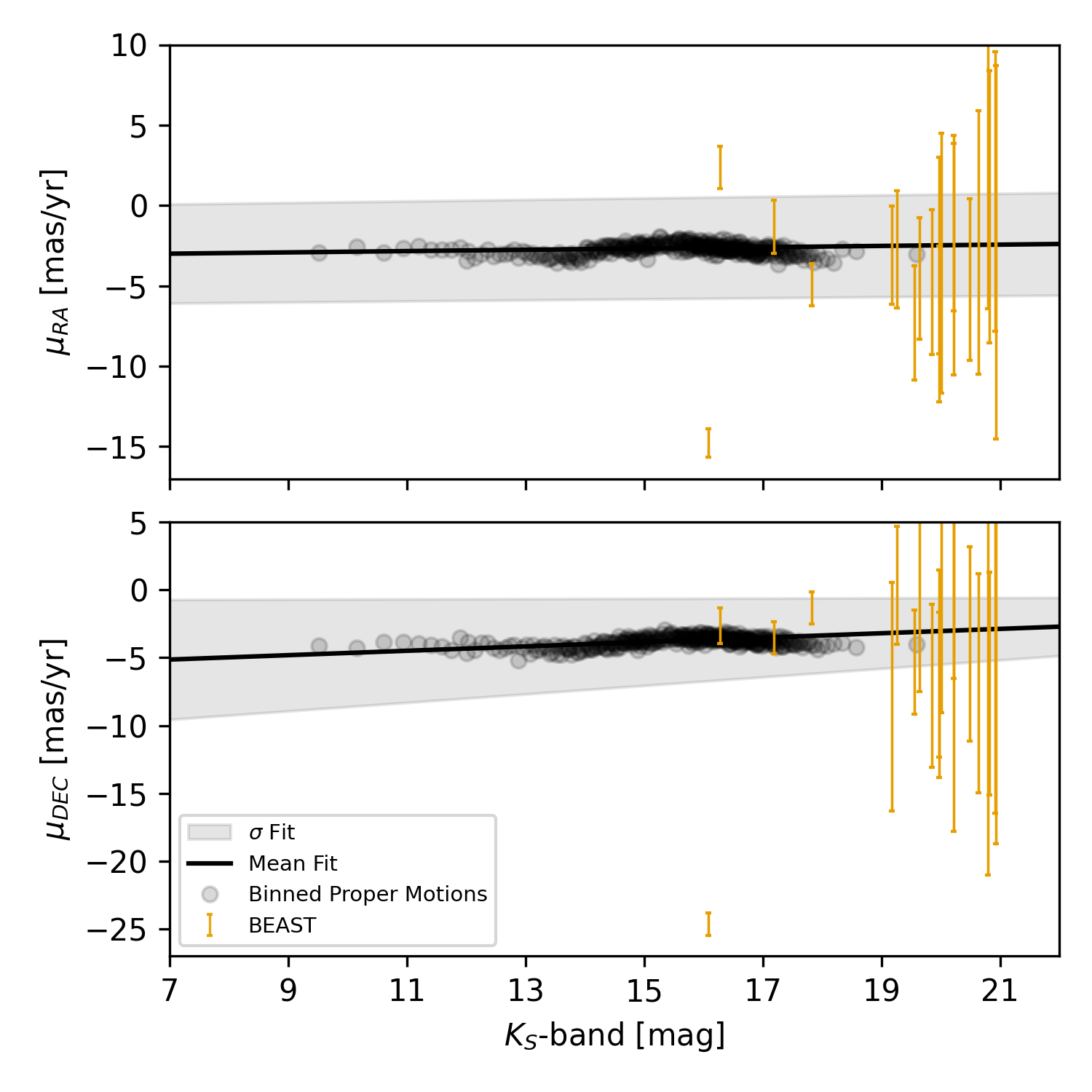}
    \caption{Proper motion model of the background stars for $\mu^2$~Sco for the right ascension direction (top) and the declination direction (bottom) in the ICRS reference system. This is similar to Fig.\ref{fig:meandist_fit}, but now includes the BEAST candidates shown in orange and transformed to ICRS.}
\label{fig:mu2Sco_pm_mag_fit}
\end{figure}

\begin{table*}
\centering
    \caption{The 20 candidates with logarithmic odds ratio greater than zero under the parallax and proper motion model, sorted by decreasing $\log_{10}r_{cb}(\mu,\varpi)$ (Eq.~\ref{eq:pm_oddsratio}). 
    Those above 1.0 we identify as likely companions (in bold face). The column "Candidate ID" contains unique identifier strings assigned to each candidate by \cite{Janson_2021} in most cases, and/or the commonly accepted companion name ("B" referring to a very probable stellar-mass companion). The full table including results for 263 candidates (including six candidates published in  \citealt{Gayathri_2023} that did not have follow-up when our list was compiled) is available online.}
    \begin{tabular}{l l l l l l l l}
    \toprule
        Host Star & Host Star & Candidate ID & $K_s$-band & $N_{\rm obs}$ & $\Delta t_{\rm obs}$ & $\log_{10}r_{cb}(\mu)$ & $\log_{10}r_{cb}(\mu, \varpi)$\\
        & & & (mag) & & (years) & & \\
        \midrule
        HIP~71865 & b~Cen & b (1oh0dfaj) & 17.17& 3 & 20.9 & 15.78 & $\mathbf{\geq300}$ \\ %
        HIP~61257 & HD~109195 & "B" (58pwu3dv) & 12.81& 3 & 2.8 & 1.81 & $\mathbf{\geq300}$ \\
        HIP~81208 & HD~149274 & B & 13.63 & 2 & 2.7 & -2.71 & \textbf{91.66}\\
        HIP~82545 & $\mu^2$~Sco & b (snojiu5b) & 16.08& 2 & 3.1 & 8.00 & \textbf{51.36} \\
        HIP~81208 & HD~149274 & C & 12.56 & 2 & 2.7 & 2.29 & \textbf{42.58} \\
        HIP~52742 & HD~93563 & x9ld1uh0 & 12.03& 2 & 1.1 & 1.57 & \textbf{12.56} \\
        HIP~76048 & HD~138221 & vslvj1zp & 14.50& 2 & 0.2 & 1.42 & \textbf{2.00} \\
        HIP~60009 & $\zeta$~Cru & rtolbhl8 & 20.49& 2 & 1.0 & 0.07 & 0.32\\
        HIP~62327 & HD~110956  & n1ik1oif & 19.64& 2 & 1.0 & 0.08 & 0.30\\
        HIP~69011 & HD~123247 & 962kt2vw & 18.16& 2 & 0.2 & 0.81 & 0.26\\
        HIP~60009 & $\zeta$~Cru & ebmx222k & 19.97& 2 & 1.0 & 0.05 & 0.18\\
        HIP~74100 & HD~133937& ldj61i8l & 19.97& 2 & 1.0 & 0.08 & 0.18\\
        HIP~52742 & HD~93563 & mm9uw3ha & 19.49& 2 & 1.1 & 0.01 & 0.16\\
        HIP~60009 & $\zeta$~Cru & qhsttxkw & 20.14& 2 & 1.0 & 0.03 & 0.15\\
        HIP~60009 & $\zeta$~Cru & n41zi1pt & 19.51& 2 & 1.0 & 0.04 & 0.15\\
        HIP~60009 & $\zeta$~Cru & 4lh805ve & 18.85& 2 & 1.0 & 0.06 & 0.14\\
        HIP~60009 & $\zeta$~Cru & a7214wz3 & 19.39& 2 & 1.0 & 0.09 & 0.12\\
        HIP~60009 & $\zeta$~Cru & 5o2rkr75 & 20.11& 2 & 1.0 & 0.05 & 0.10\\
        HIP~60009 & $\zeta$~Cru & d9oco7ww & 19.01& 2 & 1.0 & 0.17 & 0.07\\
        HIP~60009 & $\zeta$~Cru & gtartoxk & 19.23& 2 & 1.0 & 0.12 & 0.06\\
        \hline
    \end{tabular}
    \label{tab:results_multipleepochs}
\end{table*}
We apply both our proper motion-only model and our proper motion and parallax covariance model to all 263 candidates. The results are shown in Table~\ref{tab:results_multipleepochs}, sorted by descending logarithmic odds ratio $\log_{10}r_{cb}(\mu, \varpi)$, i.e.\ the most likely real companions appear first. 
In what follows we focus on the results using the parallax and proper motion model in the final column.
Visualizations of all 20 candidates are shown in Fig.\ref{fig:HIP81208_B}, Fig.\ref{fig:HIP81208_C}, Figs.~\ref{fig:pm_plx_results0}, \ref{fig:pm_plx_results00} and~\ref{fig:pm_plx_results000}.
The complete table of candidate astrometry and photometry from the BEAST survey will be published in Delorme et al.\ (in preparation).

Most interesting are the seven candidates that have 
an odds ratio greater than ten (i.e.\ $\log_{10}r_{cb}(\mu, \varpi)>1$). 
Two of these, b~Cen~b and $\mu^2$~Sco~b, are well-established as exoplanets (\citealt{Janson_2021,Squicciarini_2022}).
We further infer that the candidates of HIP~81208 (two candidates), HIP~61257, HIP~52742, and HIP~76048 are likely true companions (but not necessarily in the exoplanet mass regime).
The remaining candidates in Table~\ref{tab:results_multipleepochs} are still formally favoured by the co-moving companion model, but not by much. These will require more epochs or other observations in order to determine their nature.

We now examine the results on some of the individual, high odds ratio candidates, which will also serve to illustrate how our method works.

\subsection{Results on individual candidates}

\begin{figure}
    \centering \includegraphics[width=\linewidth]{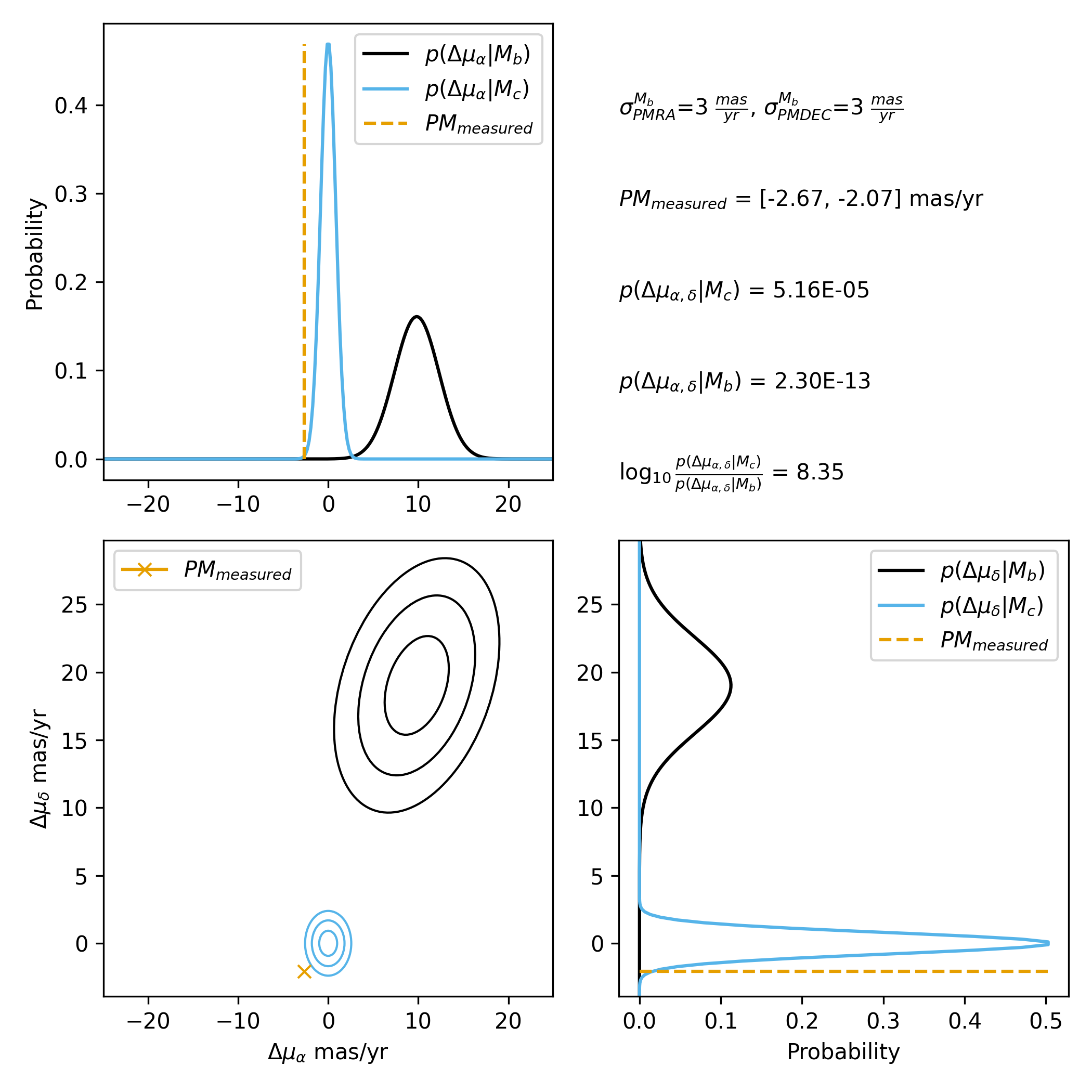}
    \caption{
    Demonstration of the proper motion-only method for the known exoplanet $\mu^2$~Sco~b. The likelihood under the background model is shown in black and the likelihood under the co-moving companion model is shown in blue. 
    The bottom left panel shows these as two-dimensional (Gaussian) distributions. The other two panels show the one-dimensional marginal distributions. The measured relative proper motion is shown in orange:
    the uncertainties in this is not shown because it is included in the two likelihoods (see Eq.~\ref{eq:pm_likleihood_truecompanion} and~\ref{eq:28}).}   
\label{fig:p_ratio_hip82545}
\end{figure}

\paragraph{$\mathbf{\mu^2}$~\textbf{Sco~b}}
The astrometric field star model for the target $\mu^2$~Sco is shown in Fig.~\ref{fig:mu2Sco_pm_mag_fit}.
The black line is our fit to the background stars (described in Section~\ref{sec:FSmodel}), which can be compared to the various BEAST candidates in this field in orange. Nearly all of the candidates agree with the field star model, showing the large degree of field star contamination that can be present in exoplanet searches. Just one does not agree; this is the exoplanet $\mu^2$~Sco~b.
The PDF over the proper motion (i.e.\ the likelihood) for this candidate under the background model is shown by the black curves/contours in Fig.~\ref{fig:p_ratio_hip82545}. The likelihood for the companion model is shown by the blue curves/contours, and the measured proper motion is shown as the orange point/line. 
In this example we see that the measurement is far more consistent with the companion model than with the background model. This is a rather clear-cut case even by visual inspection; many other cases in appendix~\ref{sec:plotsall} are more ambiguous.

\begin{figure}
    \centering
\includegraphics[width=\linewidth]{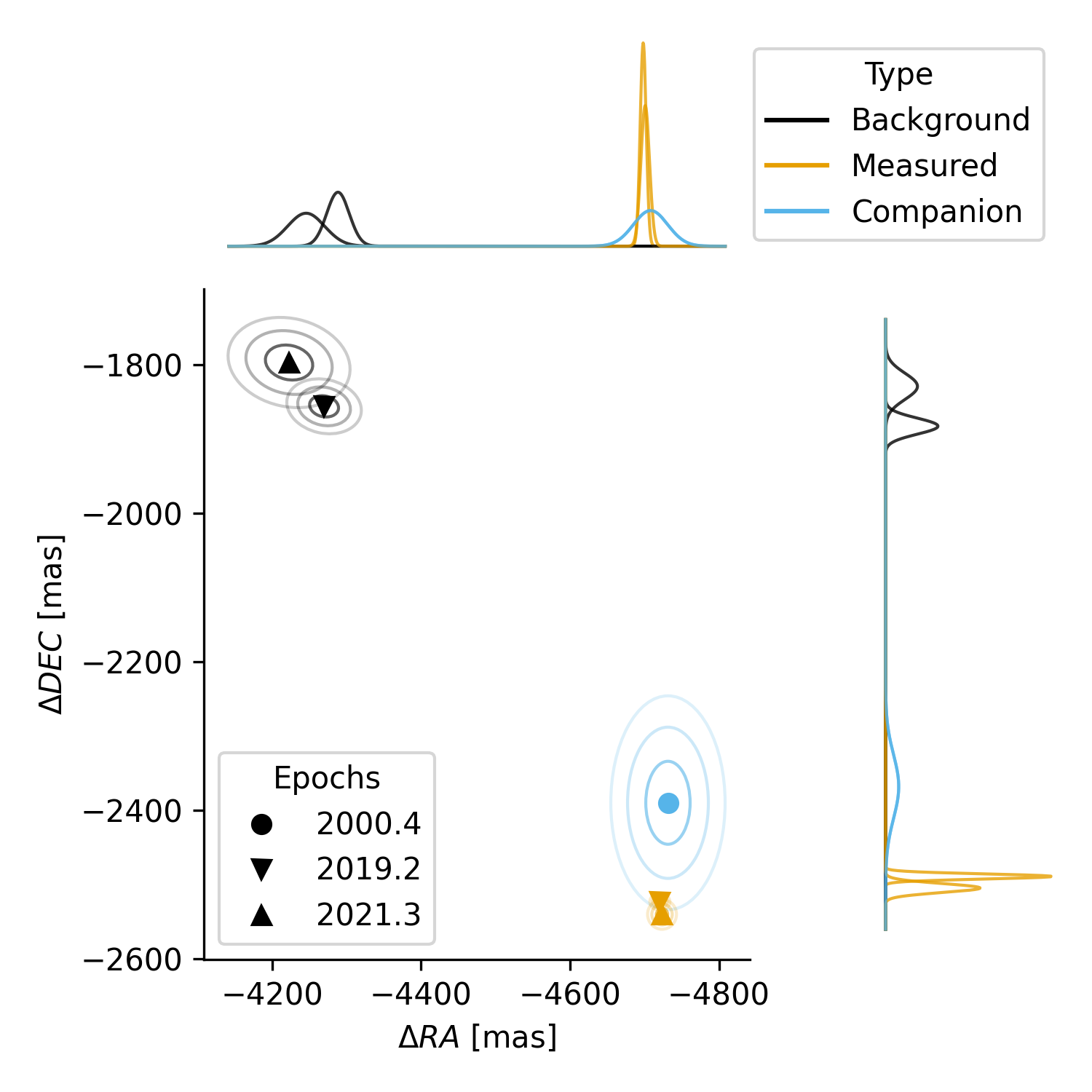}
    \caption{Visualization of the predicted positions of the candidate companion b 
    of the star HIP~71865 (b~Cen) under the proper motion and parallax model.  A co-moving companion would remain at the position of the first epoch (blue circle) because orbital motion is not included. A field star with the modelled proper motion and parallax of nearby (mostly background) stars would be measured at the two later epochs at the two positions shown by the black triangles. 
    The actual measured change in positions of the candidate are shown as orange triangles. The fact that these are much nearer to the blue distribution means this is likely to be a true companion, something that is properly quantified by our method.
    The contour lines show $50\%$, $90\%$, and $99\%$ of the enclosed probability, reflecting the propagated uncertainty in the parallaxes, proper motions, and BEAST position measurements. The marginal likelihoods are shown on both axes.
    This visualization does not show the covariances between the measurements at different epochs, which are nonetheless taken into account by our method (see Eq.~\ref{eq:Lambda_prime}). 
    }
    \label{fig:bCen_pm_plx_model}
\end{figure}

\paragraph{\textbf{b~Cen~b}}
This candidate, which was identified as an exoplanet by \cite{Janson_bCen_2021}, has three epochs in our analysis. Figure~\ref{fig:bCen_pm_plx_model} shows 
the change in positions of the candidate under the background model $M_b$ (black), the (unchanged) position of the candidate under the co-moving companion model $M_{c}$ (blue), and the measured positions of the candidate (orange). 
The first observation of this candidate was at the epoch $2000.4$ (circle) based on archival images by \cite{Shatsky2002}. In the relative reference frame to the host star, a co-moving companion would not move from its $2000.4$ position in our model, and so is not shown for other epochs.
The next observation of the candidate took place in $2019.2$ with BEAST. Based on the background proper motion and parallax distribution, as well as the host star's proper motion, a background object in the same area of the sky as the host star would have moved about $850\,\si{mas}$ since $2000.4$. However, the figure shows that by $2019.2$ the candidate has moved much less (orange point). Another epoch at $2021.3$ confirms that the motion of the candidate is much closer to the host star's motion than it is to that of the background population. The background star hypothesis is clearly ruled out in this case. 
But because the measured positions lie near the 99\% contour of our co-moving companion model, 
some residual motion compared to the co-moving companion model likely remains. This can be explained by orbital motion, which is not taken into account, over the 21-year observational baseline \citep{Janson_bCen_2021}.

\paragraph{\textbf{HIP~61257~"B"}}
This highly probable candidate of HIP~61257 can be further examined by including non-BEAST data.
\begin{figure}[t]
    \centering
    \includegraphics[width=\linewidth]{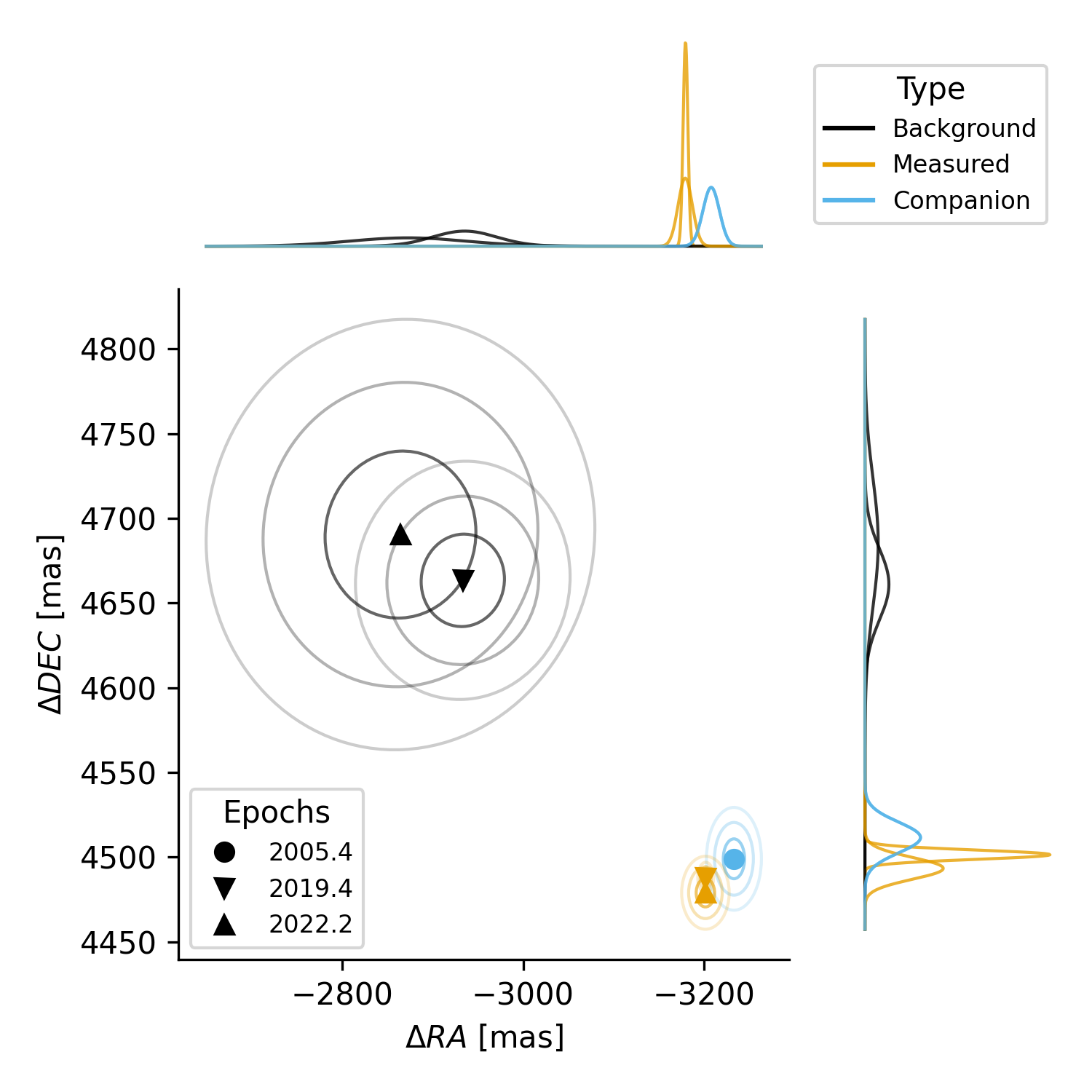}
    \caption{The results of our model applied to the  position measurements of HIP~61257 "B" from \cite{Kouwenhoven_2005} and \cite{Janson_2021} over a 17-year baseline. 
    See the caption to Figure~\ref{fig:bCen_pm_plx_model} for a description.}
    \label{fig:HIP61257_58pwu3dv}
\end{figure}
\cite{Kouwenhoven_2005} discuss a potential companion around HIP~61257, but they ultimately classify it as a background object. This potential companion has a $K_S$-band magnitude of $12.43$\,mag and a separation of $5540\,\si{mas}$, which coincides with the magnitude and separation of one of our candidates with logarithmic odds ratio greater than zero in both modelling frameworks. The astrometric motion plot (Fig.~\ref{fig:HIP61257_58pwu3dv}) shows that this candidate is co-moving with the host star. As the astrometric data from \cite{Kouwenhoven_2005} are reported without separation and position angle uncertainties, we adopted an uncertainty of $10\si{mas}$ in the relative right ascension and declination. This companion was also discussed by \cite{gratton_2023}, who identified it to be a low-mass star ($0.083 \pm 0.01 M_{\odot}$) based on its Gaia and K-band magnitude. The logarithmic odds ratios without the archival data are $\log r_{cb}(\mu, \varpi)=12.91$ and	$\log r_{cb}(\mu)=2.45$. We therefore confirm this candidate as a bona-fide binary companion based on its astrometry.

\begin{figure*}
    \centering
    
    \begin{subfigure}{0.45\textwidth}
        \includegraphics[width=\linewidth]{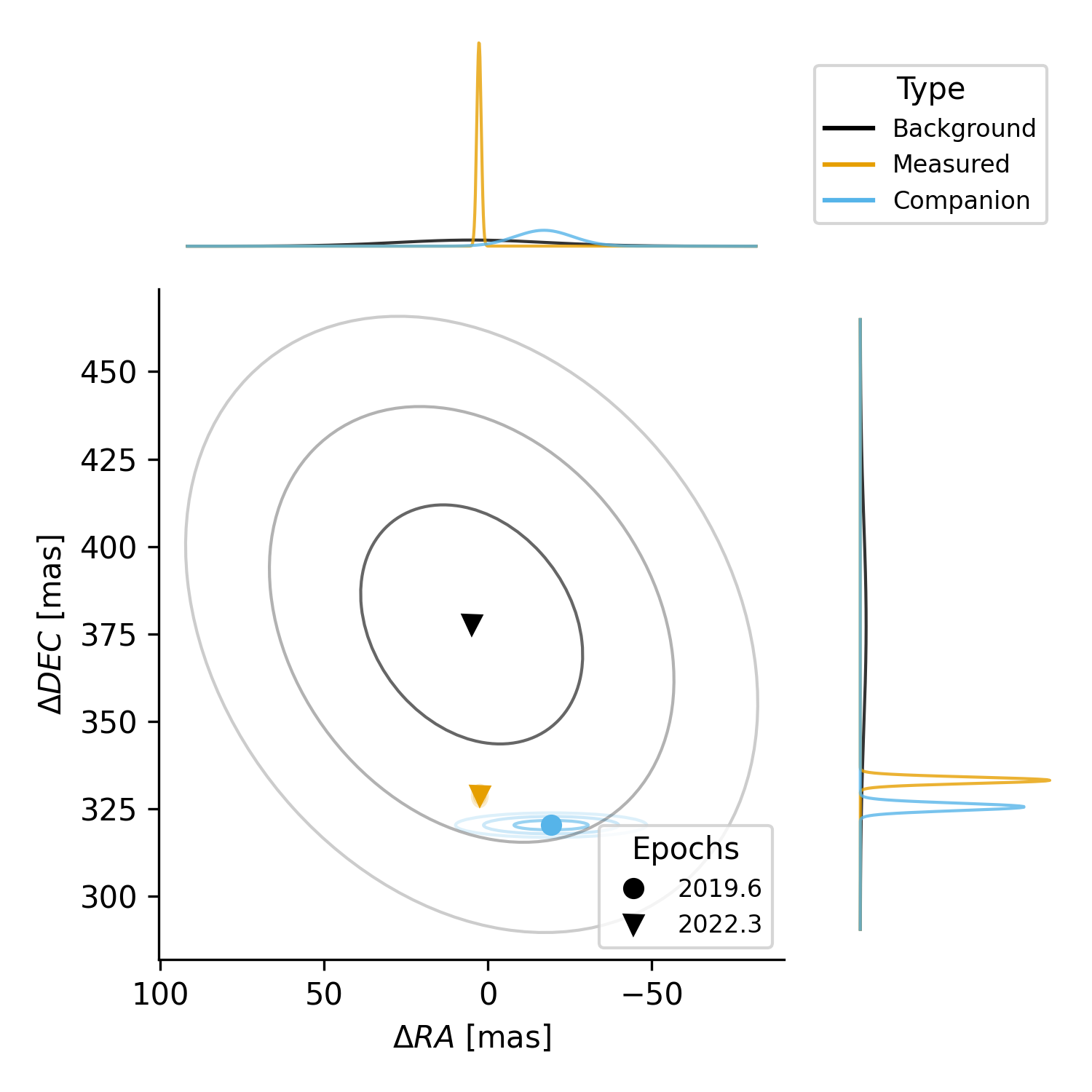}
        \caption{HIP~81208~B}
        \label{fig:HIP81208_B}
    \end{subfigure}
    \begin{subfigure}{0.45\textwidth}
        \includegraphics[width=\linewidth]{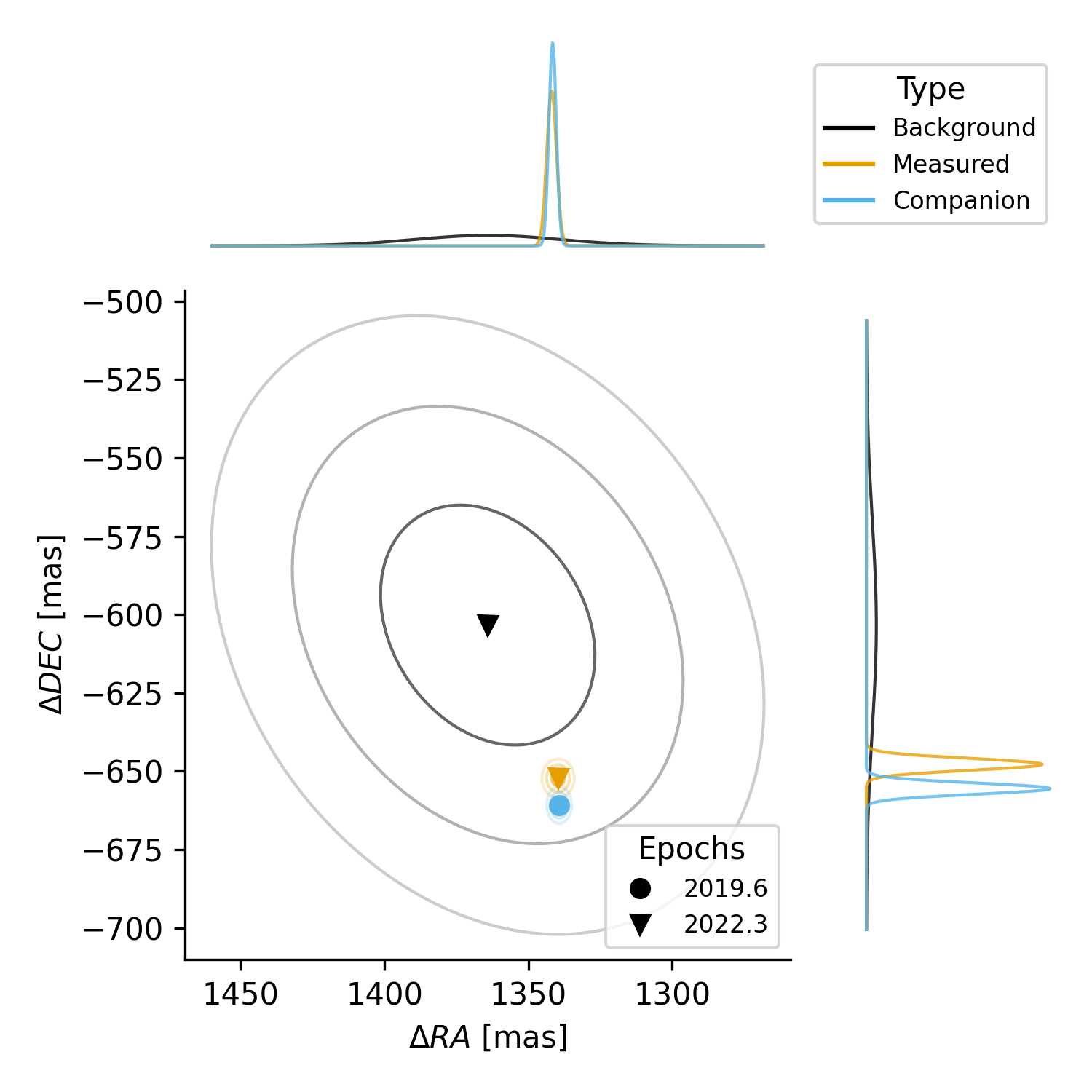}
        \caption{HIP~81208~C}
        \label{fig:HIP81208_C}
    \end{subfigure}
    \caption{Astrometric motion of the two candidates of HIP~81208 with an observation baseline of three years. \cite{Gayathri_2023} report both objects as co-moving companions based on their proper motion analysis. Our analysis supports this claim.}
    \label{fig:HIP81208}
\end{figure*}

\paragraph{\textbf{HIP~81208~B~and~C}}
These two candidates are identified as co-moving companions by \cite{Gayathri_2023}. 
They identify HIP~81208~B as a $67\,\si{MJup}$ object, mostly likely a brown dwarf, and HIP~81208~C as a $0.135\,M_{\odot}$ low-mass star. Our results for these are shown in Figs.~\ref{fig:HIP81208_B} and~\ref{fig:HIP81208_C} \citep[using astrometry and photometry from Table~C.1 in][]{Gayathri_2023}. Both candidates are favoured by the co-moving companion model with our more sophisticated parallax and proper motion method. The proper motion only model rejects HIP~81208~B but accepts HIP~81208~C. This system was recently analyzed further and found to be a gravitationally-bound hierarchical quadruple system comprised of low mass objects with a newly discovered companion to the C component \citep[Cb,][]{Chomez2023}.

\paragraph{\textbf{HIP~52742~x9ld1uh0~and~mm9uw3ha}}
Two candidates with magnitudes of $K_S=12.03$\,mag and $K_S=19.49$\,mag are formally favoured by our companion model.
The brighter one we identify as a companion with high significance (log odds ratio 12.6).
\cite{gratton_2023} 
also identified this as a companion: Adopting an age of 82.5\,Myr (obtained from assuming membership of a comoving group outside Sco-Cen; \citealt{Janson_2021}) they estimated a  mass of 0.51\,M$_\odot$ at a projected separation of 176\,au. 
In a search for astrometric acceleration from a comparison of Hipparcos and Gaia DR3 proper motions, \cite{2021ApJS..254...42B} found marginally significant evidence for an acceleration of this star, which might be evidence of this companion.
The fainter candidate companion we identify to this star has a log odds ratio of just 0.16, which is not significant.

\paragraph{\textbf{HIP~76048~vslvj1zp}}
We identify one potential companion to this star with an odds ratio of 100 in our analysis, albeit with a very short baseline of just 0.2 years.
\cite{2021ApJS..254...42B} found no significant evidence of acceleration from their Hipparcos--Gaia DR3 proper motion study.

    \section{Discussion}\label{sec:discussion}
\subsection{Interpretation}

Our analysis favours the co-moving companion model for five candidates from the BEAST data set accessible for this study, as well as two candidates in the HIP~81208 system that have a second epoch from \cite{Gayathri_2023}.
Of these seven candidates, two are confirmed exoplanets: b~Cen~AB~b \citep{Janson_bCen_2021} and $\mu^2$~Sco~b \citep{Squicciarini_2022}. HIP~61257~"B" is very likely a stellar-mass companion \citep{gratton_2023}. The two candidates of HIP~81208 are discussed as being a brown dwarf and a low-mass stellar companion by \cite{Gayathri_2023}. The remaining two companions of the host stars HIP~52742 and HIP~76048 -- with a measurement baseline of only 0.2 years -- are unconfirmed at the time of writing. 

Most of the targets in our analysis have only two observation epochs with small temporal baselines (of order one year), so many candidates show little  motion relative to the host star between the epochs. It is just these cases where ad hoc approaches to assessing companionship are inconclusive, and our statistical framework is most useful.
For many of these short baselines the co-moving model is not favoured. Longer temporal baselines make it easier to distinguish between the models, especially as the targets of direct imaging campaigns tend to be nearby stars with large proper motion. 

\subsection{Model assumptions}

Our model does not currently take into account orbital motion, so the co-moving companion model may not be favoured even if the candidate is a true companion with significant orbital motion over the observational baseline. 
Orbital periods for directly-imaged planets tend to be long, on the order of centuries or even millennia. In those cases, neglecting the orbital motion will often not affect our analysis.
In contrast, planets with relatively short orbits like $\beta$~Pic~b and~c \citep[23.6 and 3.3 years respectively,
][]{betaPicb_orbit} can show significant orbital motion. 

If orbital motion can be directly observed then the candidate is very likely to be confirmed as a companion \citep[e.g.][]{Marois2008}.
This would be a relatively straight forward extension to our modelling approach by including a model with path curvature.
Plausible paths (priors) could be generated from a Keplerian model 
\citep[e.g.\ using][]{orbitize} and then marginalizing over them to determine the model likelihood. 
An alternative approach is to show that a candidate's motion is consistent with that of field stars while also showing that the range of orbits that can explain the data are beyond the escape velocity of the system \citep{Nielsen2017, Wagner2022}. 


We build our field star astrometric model using Gaia parallaxes and proper motions. On account of the limited depth of Gaia, we then have to extrapolate our model to the fainter magnitudes of our candidates. This can introduce biases, and thus incorrectly favour or disfavour the background model. 

\begin{figure}
    \centering
    \includegraphics[width=\linewidth]{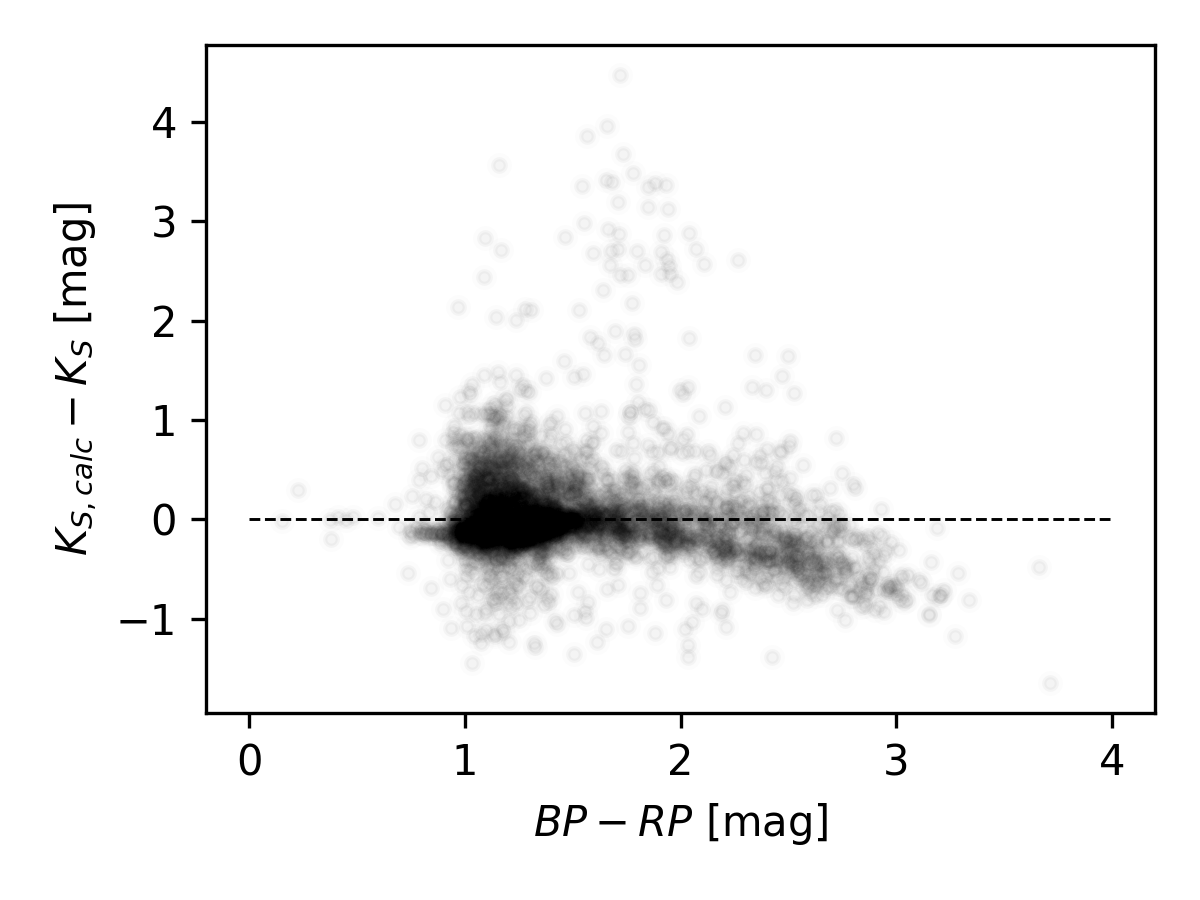}
    \caption{Difference between the $K_S$ magnitude calculated from Gaia via colour transformations and the and measured 2MASS $K_S$ magnitudes for 934\,023 objects from Gaia with 2MASS counterparts in the vicinity of $\mu^2$~Sco.}
    \label{fig:kband_diff}
\end{figure}
Another drawback of Gaia is that it observes in the optical, whereas direct imaging surveys for planets are currently done mostly in the near-infrared. To obtain the necessary infrared magnitudes of the Gaia sources we had to crossmatch to an infrared survey. We chose 2MASS because it is all-sky. However, it is not very deep, so for many field stars we instead had to predict the $K_S$-band magnitude using colour transformations from Gaia.
We used the transformations of \cite{Riello_2021} for objects in the colour range $-0.5<G_{BP}-G_{RP}<2.5$. 
The accuracy of this for brighter stars
around $\mu^2$~Sco is demonstrated in Fig.~\ref{fig:kband_diff}.
The 2MASS objects only cover a $K_S$-band range from 10 to 18\,mag while the transformed photometry extends to 21\,mag.  Some exoplanet surveys are also conducted at shorter wavelengths, such as $J$ and $H$, for which colour transformations are more reliable as they are closer to the observed Gaia bands. Ideal for our purposes, of course, would be an infrared version of Gaia.

As explained in section~\ref{sec:FSmodel}, in our background model the parallax and proper motion distributions depend only on direction and magnitude. A further improvement would be to add dependence on the colour, if that is provided by the imaging survey. 
We could then also use the measured colour to infer something about the intrinsic properties of the object, e.g.\ its spectral type \citep[e.g.][]{Parviainen2019}. While this may help determine whether or not it is a low-mass object, it would not tell us whether it is gravitationally bound.

In computing the odds ratio we only consider kinematics. We do not take into account the number density of background stars or the angular separation between the candidate and target star. Yet for a given separation, the less dense the background, the more likely the candidate is a genuine companion \citep[e.g.][]{TAMURA2016, Squicciarini_2022}. This information could be incorporated as an additional multiplicative odds ratio, although it requires a model (or measurement) of the stellar density at the faint magnitudes at which we conduct our exoplanet survey.

Finally, we emphasise that we report our results as a ratio of likelihoods of two models, where each likelihood is the probability of the data given the model. To convert each likelihood into a posterior probability of the model given the data, we would need to adopt a prior probability for the model, and know that our models are exhaustive. The latter is not yet the case, as we have neglected orbital motion, for example. The prior could incorporate the direction-dependent variation of the background star number density.

    \section{Summary}
This work has introduced a statistical method that uses multi-epoch astrometry of an imaged exoplanet candidate to compare a co-moving companion model with a chance-aligned field star model. It puts what is commonly referred to as the ``common proper motion test''  on a probabilistic footing.

Our statistical model enables a quantitative analysis of an arbitrary number of epochs, a task that cannot be achieved effectively through visual inspection. We consider the proper motion and parallax of the host star and the candidate and evaluate the likelihoods under two different models for the candidate: one in which it is a co-moving companion with negligible orbital motion, the other in which it is a member of the field star population. For the latter we build a probabilistic model of the distribution of the proper motions and parallaxes of field stars as a function of magnitude, using a fit to Gaia data in the field of each target star. 

We applied our method to a sample of 263 candidates around 23 stars from the B-Star Exoplanet Abundance Study (BEAST).
We first developed a purely proper motion based method, which we then extended to take into account the parallax. 
This model accommodates the covariance in the astrometry both between the measurements and across multiple epochs, for both Gaia astrometry and the direct measurements.
We identify seven candidates as co-moving companions.
Five of these have been identified as real companions in the literature, including the two exoplanets $\mu^2$~Sco~b and b~Cen(AB)~b.
The remaining two candidates are priority targets for further investigation. 

Our modelling approach is publicly available as an open-source Python package on \href{https://github.com/herzphi/compass}{GitHub}, allowing for easy evaluation and visualization of existing and new data. While this work presents an improvement over current practices, there is scope for further improvement.  Of particular importance is the inclusion of exoplanet orbital motion in the companion model, the incorporation of stellar number densities, and discriminating field stars from exoplanets based on their spectral information.

\begin{acknowledgements}
We are grateful for the helpful comments during the refereeing process which improve the paper. The observations this study is based on were acquired at the ESO VLT telescope (program 1101.C-025). This publication makes use of data from the European Space Agency (ESA) mission Gaia (\url{http://www.cosmos.esa.int/gaia}), processed by the Gaia Data Processing and Analysis Consortium (DPAC, \url{http://www.cosmos.esa.int/web/gaia/dpac/consortium}). Funding for the DPAC has been provided by national institutions, in particular the institutions participating in the Gaia Multilateral Agreement. 
This publication also makes use of data from the Two Micron All Sky Survey, which is a joint project of the University of Massachusetts and the Infrared Processing and Analysis Center/California Institute of Technology, funded by the National Aeronautics and Space Administration and the National Science Foundation.
\end{acknowledgements}

    \bibliographystyle{bibtex/aa} 
    \bibliography{references.bib} 
    \appendix\section{Mathematical description of the multi-epoch probabilistic model}\label{sec:2Ndmodel}

We describe here in more detail the probabilistic model introduced in section~\ref{sec:overall_prob_model}.
In high-contrast imaging, the set of positions of candidates are generally determined relative to the host star, i.e.\
\begin{equation}\label{eq:rel_position}
    \begin{split}
        \Delta \boldsymbol{x}' &= \boldsymbol{x}_{a}'-\boldsymbol{x}_{\star}'\\
        \Delta \boldsymbol{y}' &= \boldsymbol{y}_{a}'-\boldsymbol{y}_{\star}'
    \end{split}
\end{equation}
where the subscript $a$ refers to the candidate and ${\star}$ to the host star. The accents on the coordinates $\boldsymbol{x}_{a}'$ and $\boldsymbol{x}_{\star}'$ indicate that they are ``true'', i.e.\ without positional measurement uncertainties. The noisy, observed versions are $\boldsymbol{x}_{a}$ and $\boldsymbol{x}_{\star}$ at the (noise-free) times $\boldsymbol{t}$. The times we use in our models are measured relative to the Gaia epoch of observation of the host star, because we use Gaia data to build our proper motion and parallax distribution models for the field stars. 

As the true positions are nonetheless predictions based on noisy parallaxes and proper motions (and propagated in time using Eq.~\ref{eq:rel_position_Nepochs}), 
we write the probability density functions of these true positions as
\begin{equation}
        P(\Delta \boldsymbol{x}', \Delta \boldsymbol{y}'\mid \boldsymbol{t}, M(\varpi, \boldsymbol{\mu}))
\end{equation}
where $M$ can be $M_{c}$ or $M_b$ and the PDF can be written as Gaussians with the variances and covariances between the two positions $\Delta \boldsymbol{x}'$ and $\Delta \boldsymbol{y}'$. 
For the companion model, these (co)variances are derived from the Gaia data for the host star. 
For the background model, these (co)variances come from our fit to the distribution of the Gaia parallaxes and proper motions of a set of field stars, as described in section~\ref{sec:FSmodel}.
The likelihoods for both of these model (which appear in the odds ratio of Eq.~\ref{eq:pm_oddsratio}) are obtained by marginalizing over the unknown true positions $\left(\Delta\boldsymbol{x}', \Delta\boldsymbol{y}'\right)$ predicted by the model, as follows
\begin{alignat}{2}
\label{eq:PxM}
    \begin{split}
        &P\left( \Delta\boldsymbol{x}, \Delta\boldsymbol{y} \mid \boldsymbol{t}, M(\varpi, \mu) \right)\\ &= \int P\left( \Delta\boldsymbol{x}, \Delta\boldsymbol{y}, \Delta\boldsymbol{x}', \Delta\boldsymbol{y}' \mid \boldsymbol{t}, M(\varpi, \mu) \right) d\Delta\boldsymbol{x}'d\Delta\boldsymbol{y}'\\
        &= \int P\left( \Delta\boldsymbol{x}, \Delta\boldsymbol{y}\mid \Delta\boldsymbol{x}', \Delta\boldsymbol{y}'\right) \,\times \\ &\hspace*{2.5em} P\left(\Delta\boldsymbol{x}', \Delta\boldsymbol{y}' \mid \boldsymbol{t}, M(\varpi, \mu) \right) d\Delta\boldsymbol{x}'d\Delta\boldsymbol{y}' \ .
    \end{split}
\end{alignat}
$P(\Delta\boldsymbol{x}, \Delta\boldsymbol{y}\mid \Delta\boldsymbol{x}', \Delta\boldsymbol{y}')$
is the likelihood of the measured positions given the true positions, and so reflects the noise in the determination of the image centroid. We take this to be a Gaussian 
with mean $(\Delta\boldsymbol{x}', \Delta\boldsymbol{y}')$ and 2$N$-dimensional Gaussian $\Lambda$, where $N$ is the number of measurement epochs.
Equation~\ref{eq:PxM} is a convolution of two Gaussians, which results in another Gaussian. 

In the \textit{background model} 
$M_b$, $P\left(\Delta\boldsymbol{x}', \Delta\boldsymbol{y}' \mid t_i, M_{b} \right)$ is a Gaussian with mean $\left(\Delta\boldsymbol{x}', \Delta\boldsymbol{y}'\right)$ and covariance matrix
\begin{equation}\label{eq:Lambda_prime}
    \tiny
    \Lambda' = \begin{bmatrix}
         \text{Var}(\Delta x'_1) & \text{Cov}(\Delta x'_1, \Delta y'_1) & \text{Cov}(\Delta x'_1, \Delta x'_2) & \text{Cov}(\Delta x'_1, \Delta y'_2) & \ldots \\
         & \text{Var}(\Delta y'_1) & \text{Cov}(\Delta y'_1, \Delta x'_2) & \text{Cov}(\Delta y'_1, \Delta y'_2) & \ldots \\
         &&\text{Var}(\Delta x'_2) & \text{Cov}(\Delta x'_2, \Delta y'_2) & \ldots \\
         &&&\text{Var}(\Delta y'_2) &  \ldots \\
         \vdots & \vdots & \vdots & \vdots & \ddots
    \end{bmatrix}
\end{equation}
where terms are given in the order $(\Delta x_1', \Delta y_1', \Delta x_2', \Delta y_2', \ldots)$.
The symmetry of the covariance matrix fills the lower triangle of Eq.~\ref{eq:Lambda_prime}. The individual terms of $\Lambda'$ are computed from Eq.~\ref{eq:rel_position_Nepochs} to be 
\begin{equation}\label{eq:var_covar_equal}
\begin{split}
    \text{Var}(\Delta x'_i) &= t_i^2\left[\text{Var}(\mu_{b,x}) + \text{Var}(\mu_{\star,x})\right]\\
    &+ s_x(t_i)^2\left[\text{Var}(\varpi_b) + \text{Var}(\varpi_\star)\right] \\
    &+ 2t_is_x(t_i)\left[\text{Cov}(\varpi_b, \mu_{b,x}) + \text{Cov}(\varpi_\star, \mu_{\star,x})\right] \\
    \text{Var}(\Delta y'_i) &= t_i^2\left[\text{Var}(\mu_{b,y})+\text{Var}(\mu_{\star,y})\right]\\
    &+s_y(t_i)^2\left[\text{Var}(\varpi_b)+\text{Var}(\varpi_\star)\right] \\
    &+ 2t_is_y(t_i)\left[\text{Cov}(\varpi_b, \mu_{b,y})+\text{Cov}(\varpi_\star, \mu_{\star,y})\right] \\
    \text{Cov}(\Delta x'_i, \Delta y'_i) &= t_i^2 \left[\text{Cov}(\mu_{b,x}, \mu_{b,y})+ \text{Cov}(\mu_{\star,x}, \mu_{\star,y})\right]\\
    &+ s_x(t_i)s_y(t_i)\left[\text{Var}(\varpi_b)+\text{Var}(\varpi_\star)\right] \\
    &+ t_is_y(t_i)\left[\text{Cov}(\varpi_b, \mu_{b,x})+\text{Cov}(\varpi_\star, \mu_{\star,x})\right]\\
    &+ t_is_x(t_i)\left[\text{Cov}(\varpi_b, \mu_{b,y})+\text{Cov}(\varpi_\star, \mu_{\star,y})\right]\\
    \text{Cov}(\Delta x'_i, \Delta x'_j) &= 
        t_it_j\left[\text{Var}(\mu_{b,x}) + \text{Var}(\mu_{\star,x})\right]\\
        &+t_is_x(t_j)\left[\text{Cov}(\varpi_b,\mu_{b,x}) + \text{Cov}(\varpi_{\star},\mu_{\star,x})\right]\\
        &+t_js_x(t_i)\left[\text{Cov}(\varpi_b,\mu_{b,x}) + \text{Cov}(\varpi_{\star},\mu_{\star,x})\right]\\
        &+s_x(t_i)s_x(t_j)\left[\text{Var}(\varpi_{b}) + \text{Var}(\varpi_{\star})\right]\\
        \text{Cov}(\Delta y'_i, \Delta y'_j) &= 
        t_it_j\left[\text{Var}(\mu_{b,y}) + \text{Var}(\mu_{\star,y})\right]\\
        &+t_is_y(t_j)\left[\text{Cov}(\varpi_b,\mu_{b,y}) + \text{Cov}(\varpi_{\star},\mu_{\star,y})\right]\\
        &+t_js_y(t_i)\left[\text{Cov}(\varpi_b,\mu_{b,y}) + \text{Cov}(\varpi_{\star},\mu_{\star,y})\right]\\
        &+s_y(t_i)s_y(t_j)\left[\text{Var}(\varpi_{b}) + \text{Var}(\varpi_{\star})\right]\\
        \text{Cov}(\Delta x'_i, \Delta y'_j) &= 
        t_it_j\left[\text{Cov}(\mu_{b,x}, \mu_{b,y}) + \text{Cov}(\mu_{\star,x}, \mu_{\star,y})\right]\\
        &+t_is_y(t_j)\left[\text{Cov}(\varpi_{b}, \mu_{b,x}) + \text{Cov}(\varpi_{\star}, \mu_{\star,x})\right]\\
        &+t_js_x(t_i)\left[\text{Cov}(\varpi_{b}, \mu_{b,y}) + \text{Cov}(\varpi_{\star}, \mu_{\star,y})\right]\\
        &+s_x(t_i)s_y(t_j)\left[\text{Var}(\varpi_{b}) + \text{Var}(\varpi_{\star})\right].\\
\end{split}
\end{equation}
From the convolution in Eq.~\ref{eq:PxM}, we see that $P\left( \Delta\boldsymbol{x},\Delta\boldsymbol{y} \mid \boldsymbol{t}, M_b \right)$ is a $2N$-dimensional Gaussian with mean $\left(\Delta\boldsymbol{x}',\Delta\boldsymbol{y}'\right)$ and covariance matrix $\Lambda + \Lambda'$.
Note that equations~\ref{eq:var_covar_equal} assume that there is no covariance between the field star proper motion and parallax and the host star proper motion and parallax. This is not strictly true, because both the background model and the host star data are drawn from Gaia, and Gaia shows correlations between sources separated by small angles on the sky. But in practice our background model describes the distribution of an ensemble of stars, and so is less affected by Gaia's correlations between individual sources.

In the \textit{co-moving companion model} $M_{c}$, the position of the companion relative to the star is constant, leading to a zero covariance for the second term under the integral in Eq.~\ref{eq:PxM} making $P\left(\Delta\boldsymbol{x}', \Delta\boldsymbol{y}' \mid \boldsymbol{t}, M_{c} \right)$ a delta function.  $P\left( \Delta\boldsymbol{x}, \Delta\boldsymbol{y} \mid \boldsymbol{t}, M_{c} \right)$ is therefore a $2N$-dimensional Gaussian with mean $\left(\Delta\boldsymbol{x}', \Delta\boldsymbol{y}'\right)$ and covariance $\Lambda$.

\section{Candidates with positive log odds ratios}\label{sec:plotsall}

\begin{figure*}[h]
    \centering
    \includegraphics[width=0.81\linewidth]{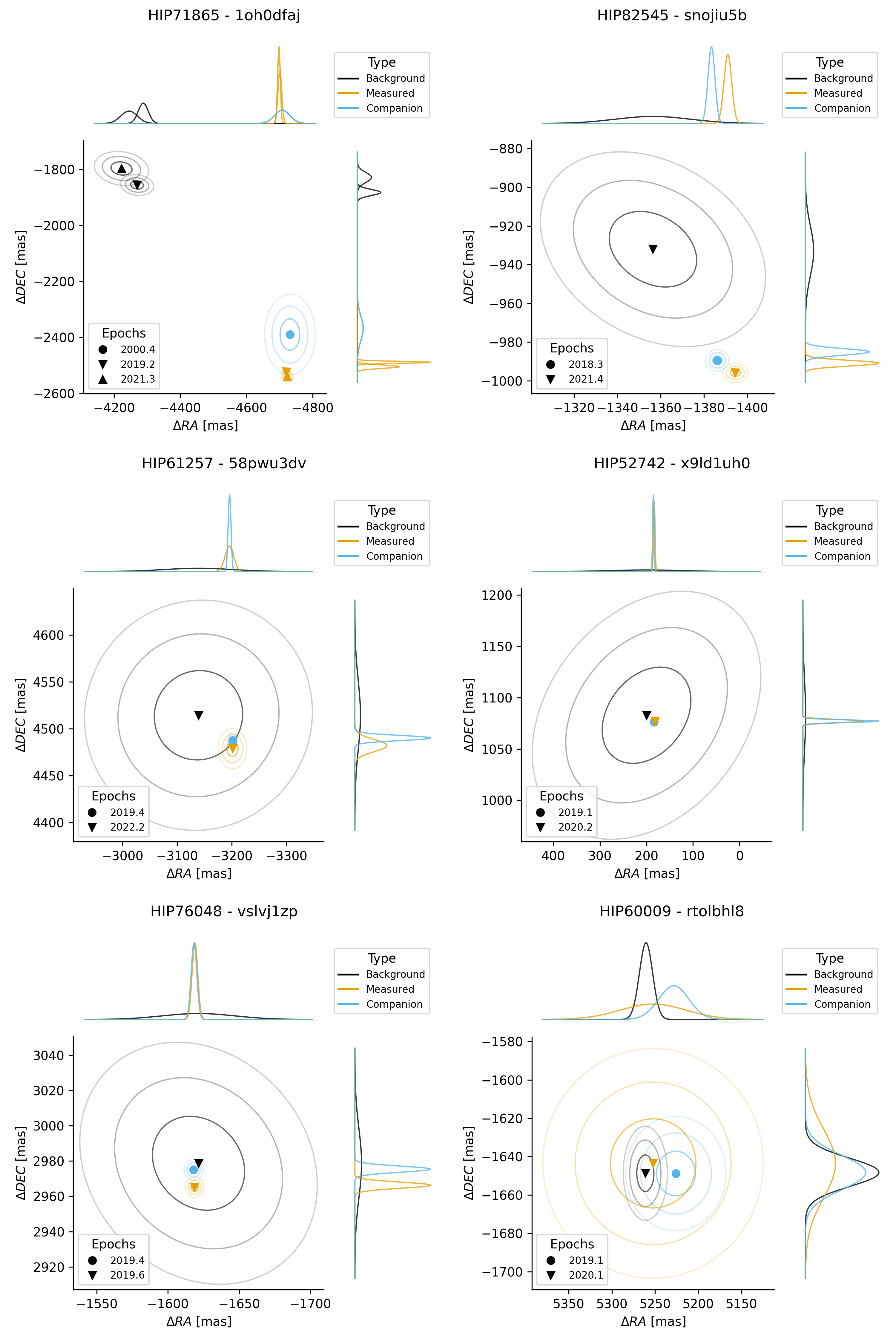}
    \caption{Candidates from BEAST with an logarithmic odds ratio $\log r_{tcb} > 0$. The schematics of the plots are explained in Figure~\ref{fig:bCen_pm_plx_model}.}
    \label{fig:pm_plx_results0}
\end{figure*}

\begin{figure*}
    \centering
    \includegraphics[width=0.88\textwidth]{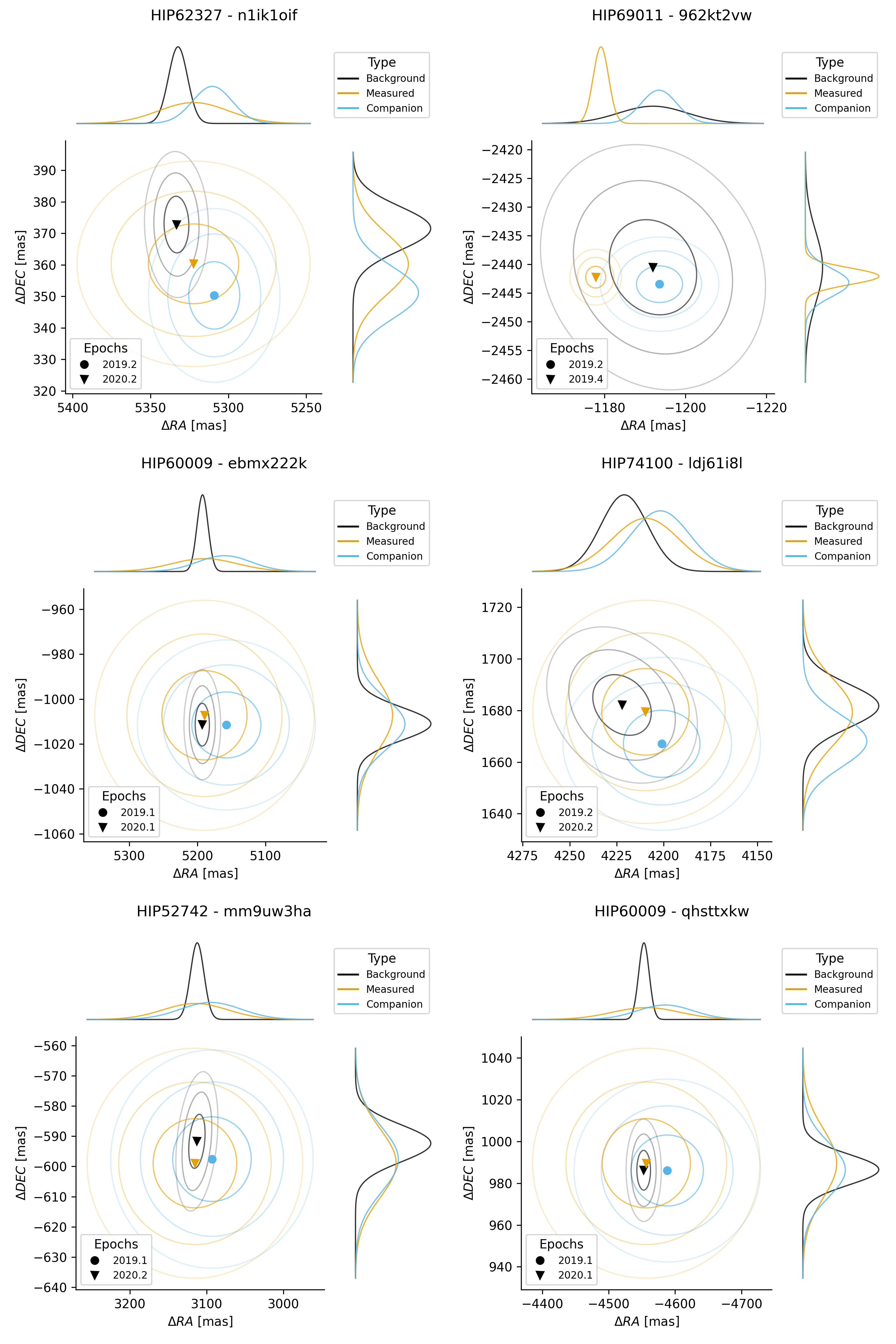}
    \caption{Candidates from BEAST with an logarithmic odds ratio $\log r_{tcb} > 0$. The schematics of the plots are explained in Figure~\ref{fig:bCen_pm_plx_model}.}
    \label{fig:pm_plx_results00}
\end{figure*}

\begin{figure*}
    \centering
    \includegraphics[width=0.88\textwidth]{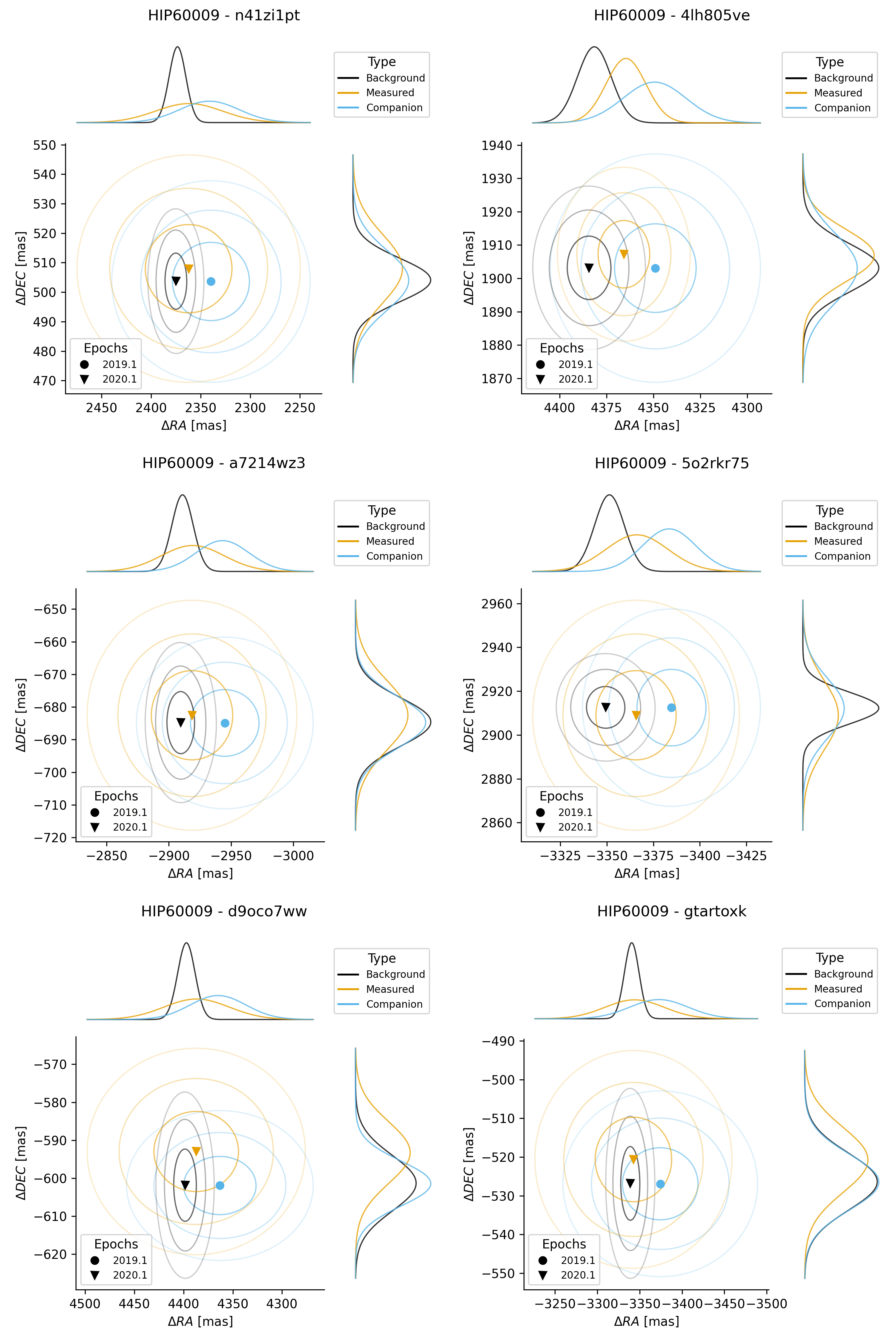}
    \caption{Candidates from BEAST with an logarithmic odds ratio $\log r_{tcb} > 0$. The schematics of the plots are explained in Figure~\ref{fig:bCen_pm_plx_model}.}
    \label{fig:pm_plx_results000}
\end{figure*}

Figures~\ref{fig:pm_plx_results0}, \ref{fig:pm_plx_results00}, and~\ref{fig:pm_plx_results000} show the modelling results for the 18 BEAST-only candidates that have logarithmic odds ratio $\log r_{tcb} > 0$. The results for the two targets with additional data are shown in Figure~\ref{fig:HIP81208}. 

\end{document}